\documentclass[useAMS,usenatbib,usegraphicx]{mn2e}
\usepackage{amssymb}
\usepackage{times}
\usepackage{aas_macros} 
\usepackage[usenames]{color}

\title[Asteroseismic measurement of surface-to-core rotation]{Asteroseismic measurement of surface-to-core rotation in a main sequence A star, KIC\,11145123 }

\author[Kurtz et al.]
{Donald W. Kurtz$^1$, Hideyuki Saio$^{2}$, Masao Takata$^3$, Hiromoto Shibahashi$^3$, \newauthor{Simon J. Murphy$^{1,4}$, Takashi Sekii$^{5}$} \\
$^{1}$Jeremiah Horrocks Institute of Astrophysics, University of Central
Lancashire, Preston PR1 2HE, UK\\
$^{2}$Astronomical Institute, Graduate School of Science, Tohoku University, Sendai, Miyagi 980-8578, Japan \\
$^{3}$Department of Astronomy, School of Science, The University of Tokyo, Bunkyo-ku, Tokyo 113-0033, Japan \\
$^{4}$Sydney Institute for Astronomy, School of Physics, The University of Sydney, NSW 2006, Australia \\
$^{5}$National Astronomical Observatory of Japan, 2-21-1 Osawa, Mitaka, Tokyo 181-8588, Japan
}

\begin{document}

\maketitle

\begin{abstract}
We have discovered rotationally split core g-mode triplets and surface p-mode triplets and quintuplets in a terminal age main sequence A star, KIC\,11145123, that shows both $\delta$\,Sct p-mode pulsations and $\gamma$\,Dor g-mode pulsations. This gives the first robust determination of the rotation of the deep core and surface of a main sequence star, essentially model-independently. We find its rotation to be nearly uniform with a period near 100\,d, but we show with high confidence that the surface rotates slightly faster than the core. A strong angular momentum transfer mechanism must be operating to produce the nearly rigid rotation, and a mechanism other than viscosity must be operating to produce a more rapidly rotating surface than core. Our asteroseismic result, along with previous asteroseismic constraints on internal rotation in some B stars, and measurements of internal rotation in some subgiant, giant and white dwarf stars, has made angular momentum transport in stars throughout their lifetimes an observational science.  
\end{abstract}


\section{Introduction}
\label{sec:1}

For four hundred years, since Galileo, we have known that the Sun rotates. As the Sun and stars evolve, their cores shrink while their outer envelopes expand and are eventually ejected, carrying away angular momentum. If stars conserved angular momentum throughout their lives, the surviving compact cores -- white dwarfs and neutron stars -- would spin much faster than is observed. Recently, Kepler Mission data have revealed core rotation in two red giant stars that is $5-10$ times faster than the surface rotation (\citealt{deheuvels2012}; \citealt{beck2012}), a contrast weaker than expected. Therefore, a strong mechanism for angular momentum transport must be acting before stars become red giants, probably in the dominant main sequence phase. Yet little is known of the internal rotation and angular momentum transport of stars.

Studies of stellar rotation and angular momentum transport are important for a full understanding of stellar evolution (\citealt{pinsonneault1997}; \citealt{tayar2013}). This affects wider studies of the chemical evolution of the universe, of galaxy formation and evolution, through its strong impact on stellar structure and evolution. It is also closely connected to the stellar dynamo process, and hence contributes to our understanding of the origin of the magnetic fields in the universe.

In asteroseismology of main sequence stars, pressure modes probe the outer layers of a star, and gravity modes probe the deep interior (\citealt{unno1989}; \citealt{aerts2010}). To find both kinds of modes in one star promises a full view of the interior, a place Arthur Eddington \citep{eddington1926} described as ``less accessible to scientific investigation than any other region of our universe.'' Since the birth of helioseismology \citep{leighton1962}, observations of internal pulsational gravity modes have been eagerly sought, without clear success \citep{appourchaux2010}. Consequently, for our Sun the interior differential rotation is known only half way down to the core \citep{schou1998}. Up to now the only other observational indications of internal rotation in main sequence stars come from model-dependent studies of two $\beta$\,Cep stars for which a single low overtone p-mode rotational dipole triplet and two or three of five possible components of the lowest overtone g-mode quadrupole quintuplet have been observed. For the $\beta$\,Cep star HD\,129929 \citet{dupret2004} found indications of internal differential rotation (see also \citealt{aerts2003}), and for another $\beta$\,Cep star, $\theta$\,Oph, \citet{briquet2007} found indications of solid-body rotation. Both of these studies place only weak, model-dependent constraints on internal rotation.

Our aim is to find main sequence stars that show both the pressure modes (p\,modes) and gravity, or buoyancy, modes (g\,modes) using the exquisitely precise Kepler Mission photometric data for the purpose of observing their interior rotation from the surface right to the core. The best candidates for this are stars known as $\delta$\,Sct -- $\gamma$\,Dor hybrids, of which there are several hundred amongst the 190\,000 stars observed by Kepler during its four-year mission \citep{uytterhoeven2011}. We discuss our first success, KIC\,11145123, in this paper.

For readers new to asteroseismology, we summarise some basic concepts used in this paper. Each eigenmode of adiabatic oscillations of spherically symmetric stars is specified by the three indices, $n$, $l$ and $m$, which are called the radial order, the spherical degree and the azimuthal order, respectively. These indices represent the structure of the eigenfunction (e.g., the radial displacement). The indices $l$ and $m$ indicate the number of surface nodes, and the number of surface nodes that are lines of longitude, respectively. Modes with $l=0$, $1$ and $2$ correspond to radial, dipolar and quadrupolar modes, respectively. We adopt the convention that positive (negative) $m$ designates prograde (retrograde) modes with respect to rotation in the inertial frame. The radial order $n$ is associated with the structure in the radial direction \citep{takata2012}. We particularly follow \citet{takata2006} for the radial order of dipolar modes. Negative values of $n$ denote the radial orders of g\,modes.

\section{Observations and frequency analysis}
\label{sec:obs}

KIC\,11145123 has a {\it Kepler} magnitude ${\rm Kp} = 13$, and is a late A star. From the Kepler Input Catalogue (KIC) revised photometry \citep{huber2014}, its effective temperature is $8050 \pm 200$\,K and its surface gravity is $\log g = 4.0 \pm 0.2$ (cgs units), showing it to be a main sequence A star. The data used for the analysis in this paper are the {\it Kepler} quarters 0 to 16 (Q0 -- Q16) long cadence (LC) data. {\it Kepler} has an orbital period about the Sun of 372.4536\,d, hence the quarters are just over 93\,d. We used the multi-scale, maximum a posteriori (msMAP) pipeline data; information on the reduction pipeline can be found in the data release notes 21\footnote{https://archive.stsci.edu/kepler/data\_release.html}. To optimise the search for exoplanet transit signals, the msMAP data pipeline removes astrophysical signals with frequencies less than 0.1\,d$^{-1}$ (or periods greater than 10\,d). None of the pulsation frequencies we analyse in this paper are near to that lower limit, but if the star has a direct rotational signal, e.g. from starspots, that will have been erased by the pipeline. Since, as we show, the rotation period is near to 100\,d, any data reduction technique will struggle to find a direct signal at this period because of its similarity to the time span or the {\it Kepler} quarterly rolls. This has no effect on our analysis.

\begin{figure}
\centering
\includegraphics[width=0.9\linewidth,angle=0]{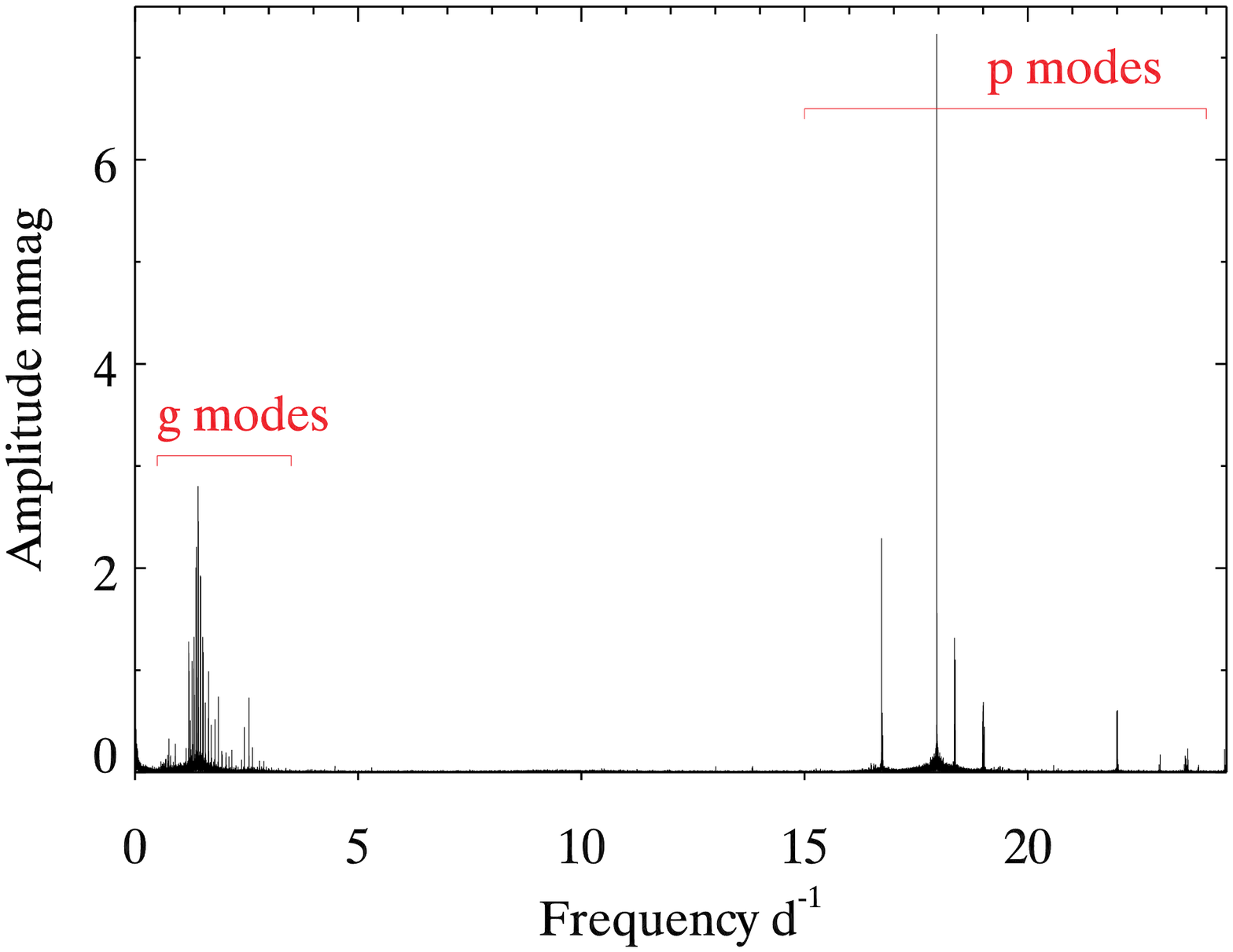}
\includegraphics[width=0.9\linewidth,angle=0]{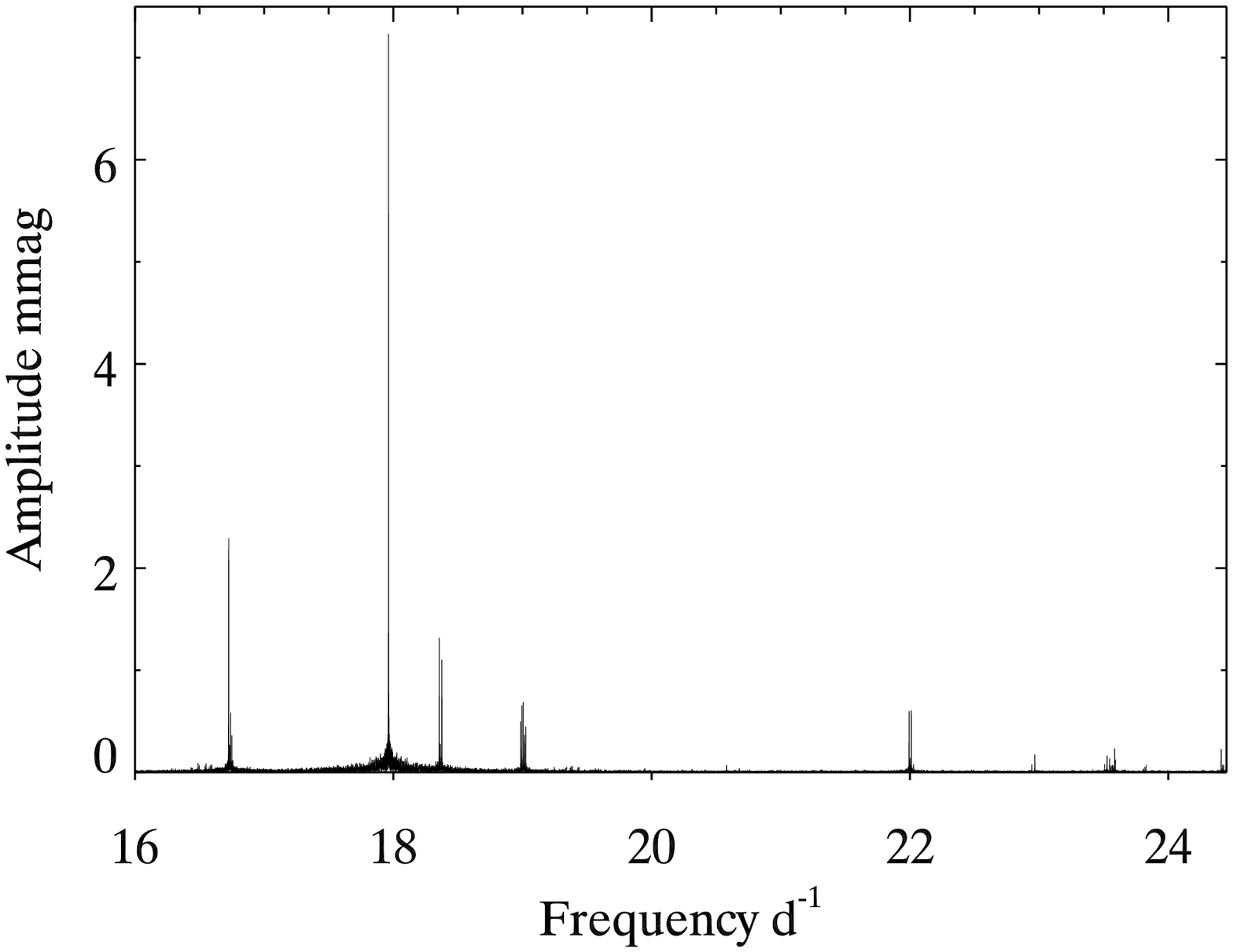}
\includegraphics[width=0.9\linewidth,angle=0]{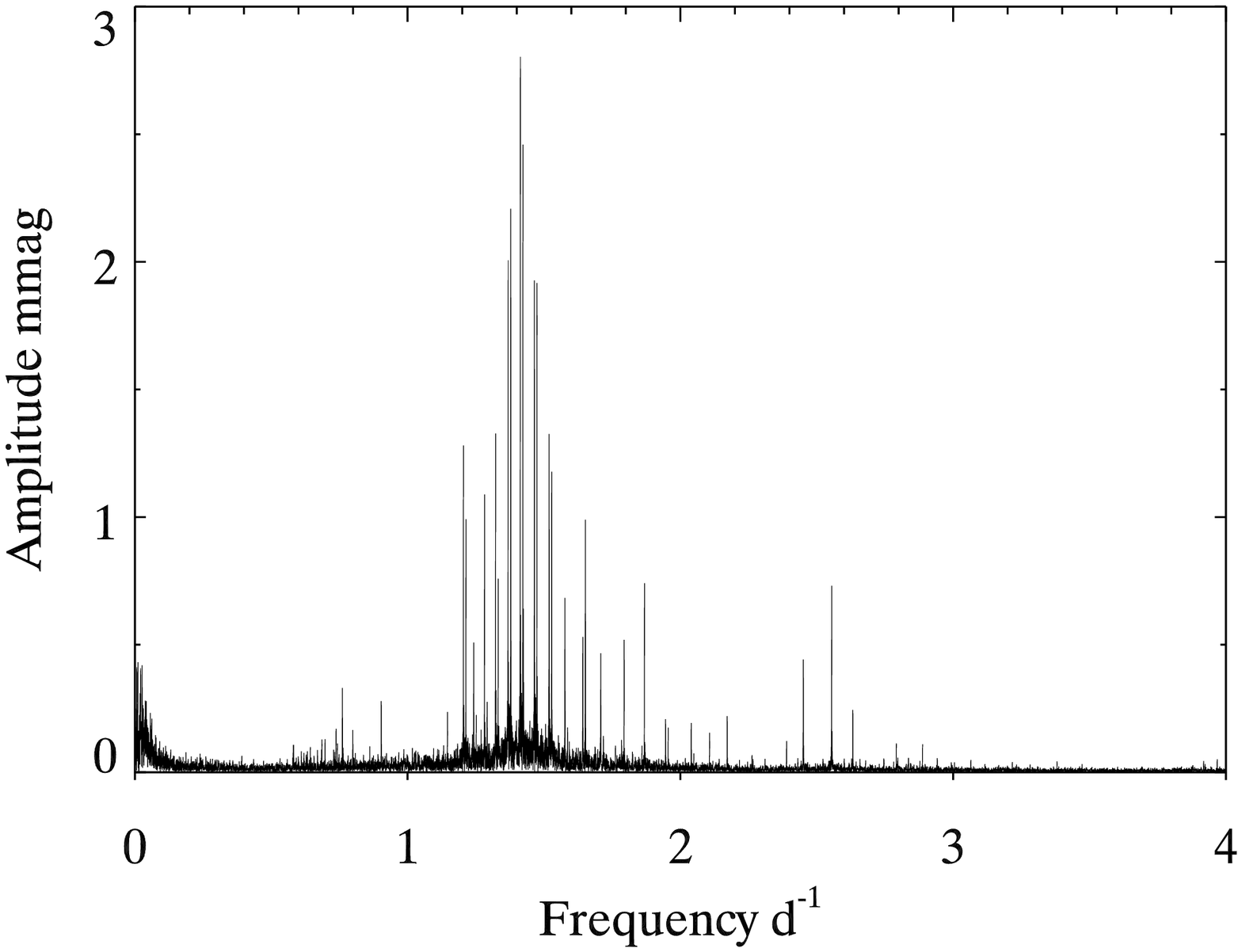}
\caption{Top panel: An amplitude spectrum for the Q0-16 {\it Kepler} long cadence data up to the Nyquist frequency for KIC\,11145123, showing the presence of both g\,modes and p\,modes that are clearly separated. The middle and bottom panels show expanded looks in the p-mode and g-mode frequency ranges, respectively.}
\label{fig:11145123_ft-all}
\end{figure}

The top panel of Fig.\,\ref{fig:11145123_ft-all} shows a full amplitude spectrum out to the Nyquist frequency for KIC\,11145123 for the nearly continuous Kepler Q0-16 LC data spanning 1340\,d (3.7\,y). There are pulsations in both the g-mode and p-mode frequency regions, which are clearly separated. The second and third panels show expanded views of those p-mode and g-mode frequency ranges, respectively.

\subsection{The p\,modes}
\label{sec:pmodes}

In the p-mode frequency range the highest amplitude peak is a singlet, which we identify as arising from a radial mode; all other higher amplitude frequencies in this range are in mode triplets, or mode quintuplets, all split by the rotation frequency in the outer envelope of the star. 

\begin{figure}
\centering
\includegraphics[width=0.9\linewidth,angle=0]{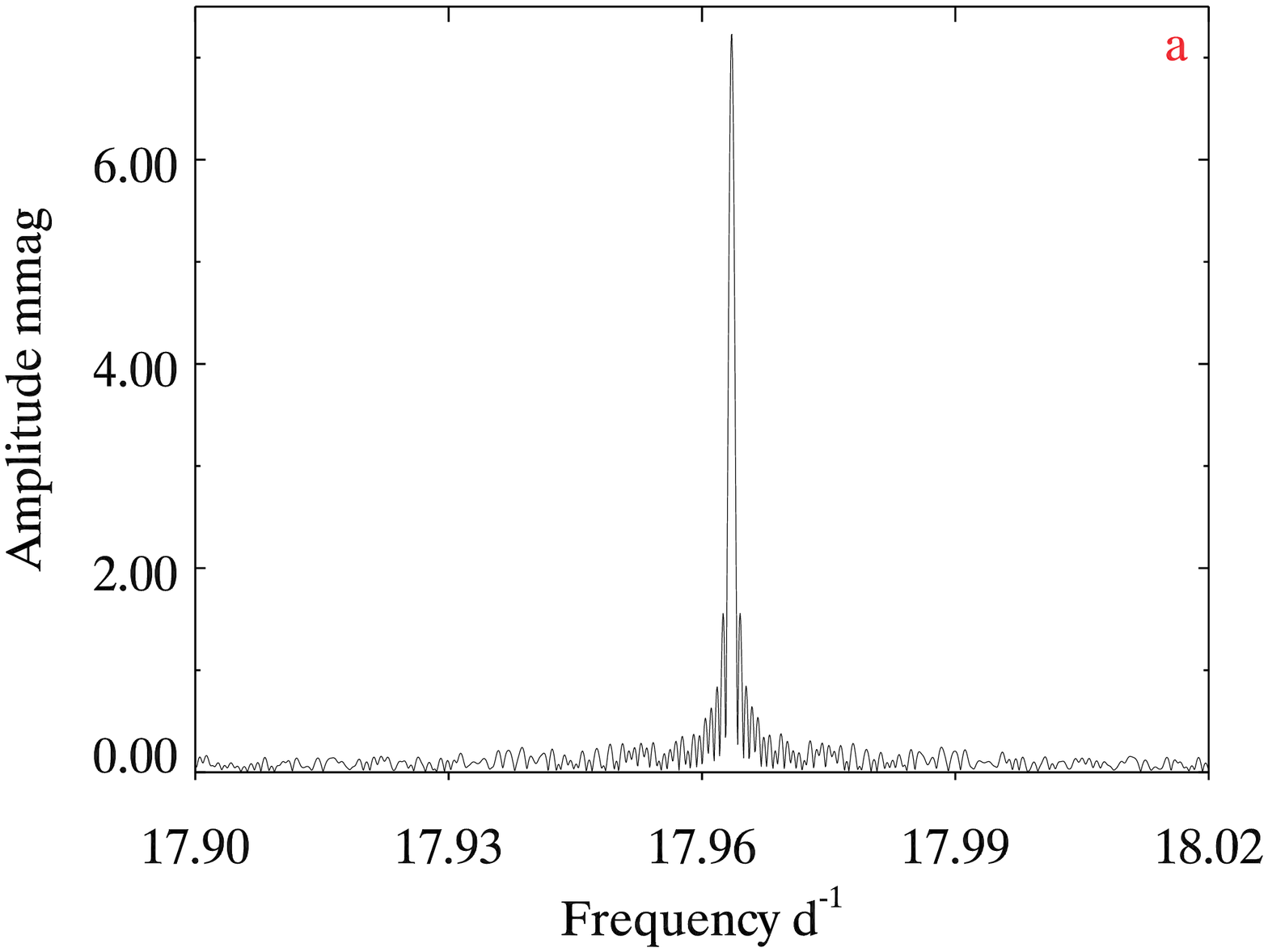}
\includegraphics[width=0.9\linewidth,angle=0]{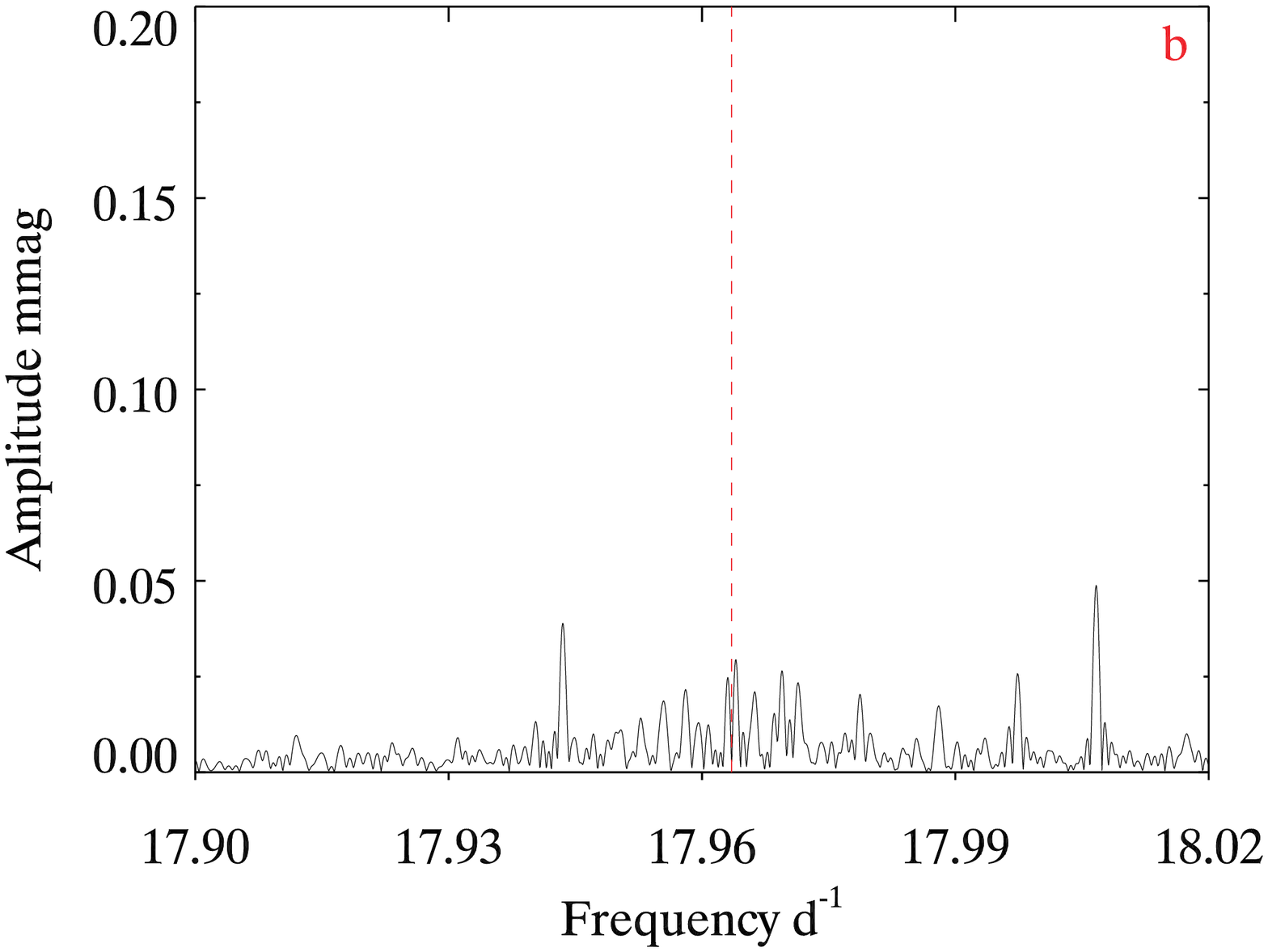}
\caption{(a): An amplitude spectrum for the highest amplitude p\,mode singlet. (b): The amplitude spectrum of the residuals after prewhitening by $\nu_1$.}
\label{fig:11145123_ft1}
\end{figure}

The highest amplitude $\delta$\,Sct p\,mode is at $\nu_1 = 17.96352$\,d$^{-1}$ and is shown in Fig.\,\ref{fig:11145123_ft1}a. For an estimate of the p-mode radial overtones, it is useful to look at the $Q$ value for $\nu_1$. This is defined to be
\begin{equation}
	Q = {P_{\rm osc}}\sqrt{\frac{\overline{\rho}}{\overline{\rho}{_{\odot}}}}
\label{eq:1}
\end{equation}
where $P_{\rm osc}$ is the pulsation period and $\overline\rho$ is the
mean density; $Q$ is known as the ``pulsation constant''.
Equation\,(\ref{eq:1}) can be rewritten as
\begin{equation}
\log Q = -6.454 +\log P_{\rm osc} +\frac{1}{2}\log g +\frac{1}{10}M_{\rm bol} +
\log T_{\rm eff},
\label{eq:2}
\end{equation}
where $P_{\rm osc}$ is given d, $\log g$ uses cgs units and $T_{\rm  eff}$ is in K. Using $T_{\rm  eff} = 8050$\,K and $\log g = 4.0$, and estimating the bolometric magnitude to be about 2 gives $Q = 0.025$, which is typical of first radial overtone pulsation in $\delta$\,Sct stars \citep{stellingwerf1979}. We thus find here only that the p\,mode frequencies are due to low overtone modes. In our best model for KIC\,11145123 discussed in Section\,\ref{sec:model} below we find the highest amplitude mode to be the second overtone radial mode.

\begin{figure*}
\centering
\includegraphics[width=0.45\linewidth,angle=0]{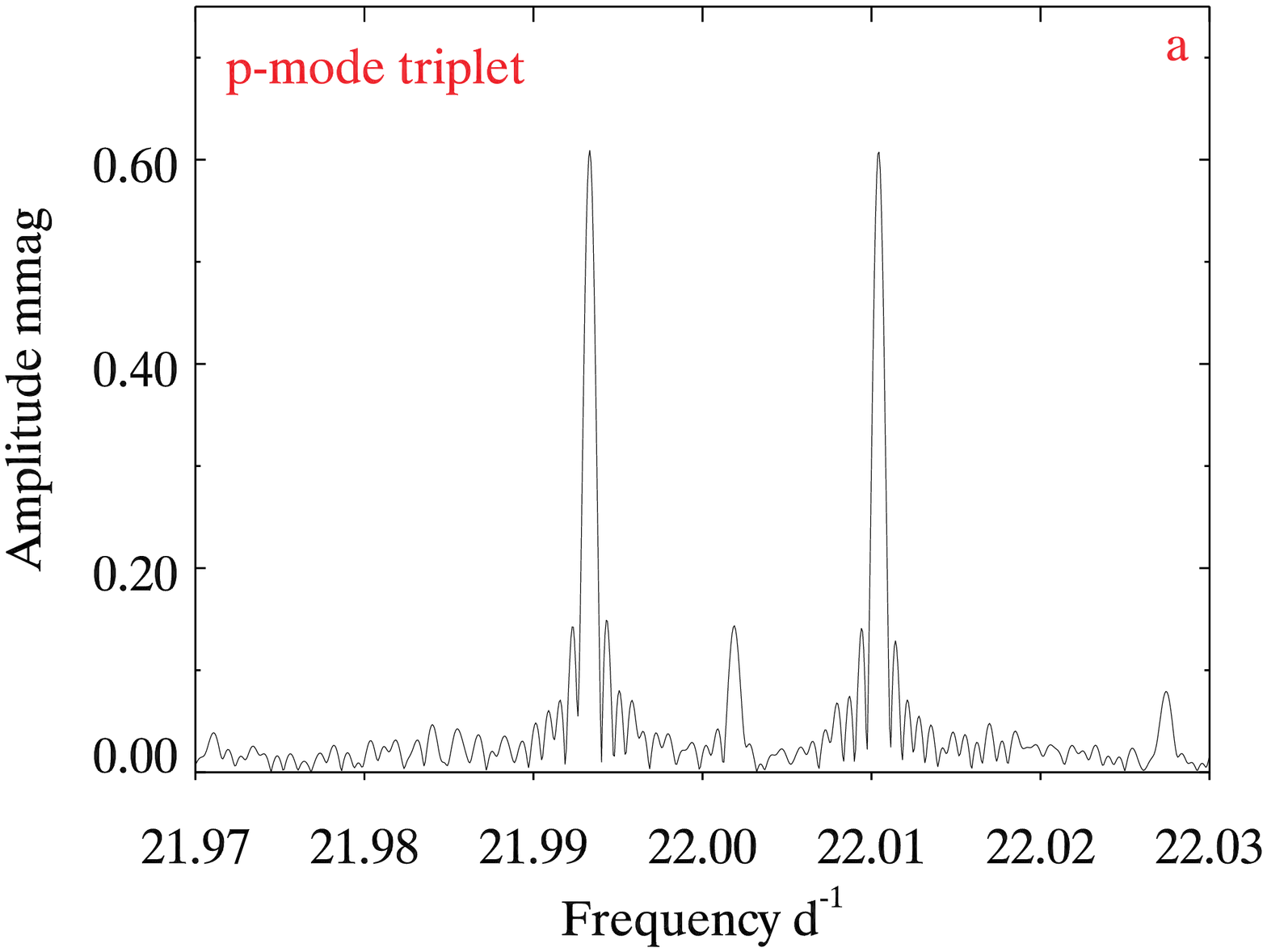}
\includegraphics[width=0.45\linewidth,angle=0]{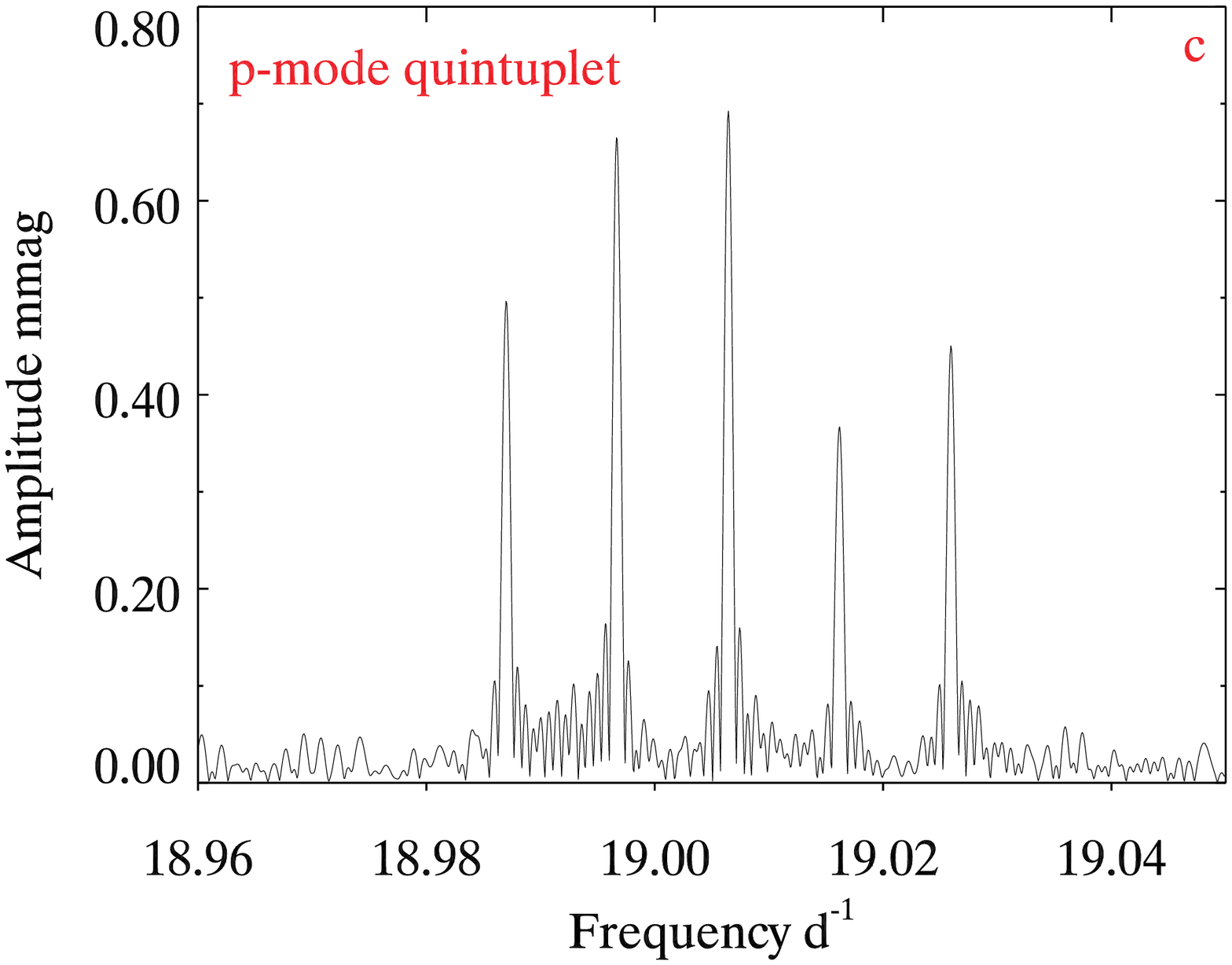}
\includegraphics[width=0.45\linewidth,angle=0]{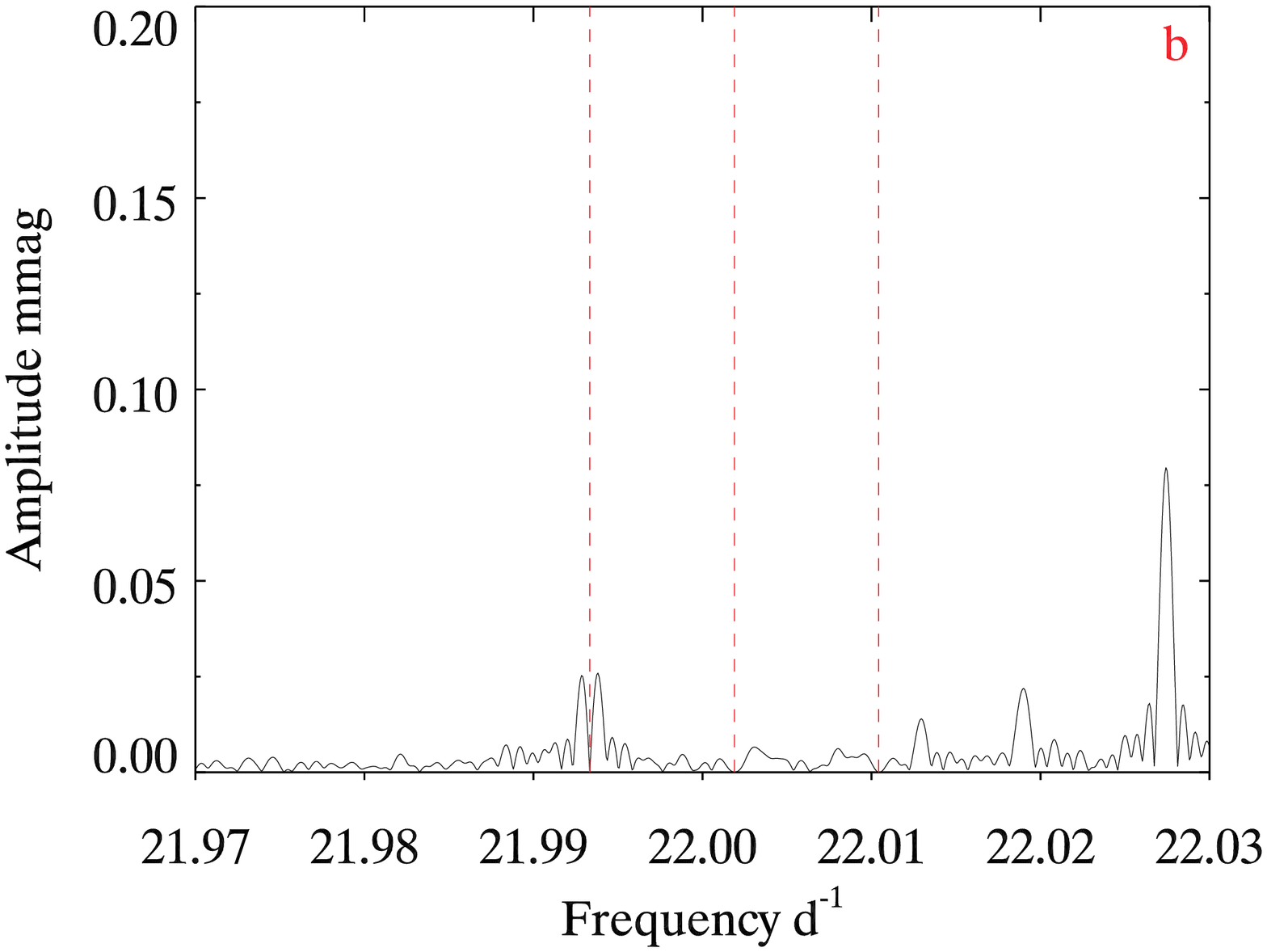}
\includegraphics[width=0.45\linewidth,angle=0]{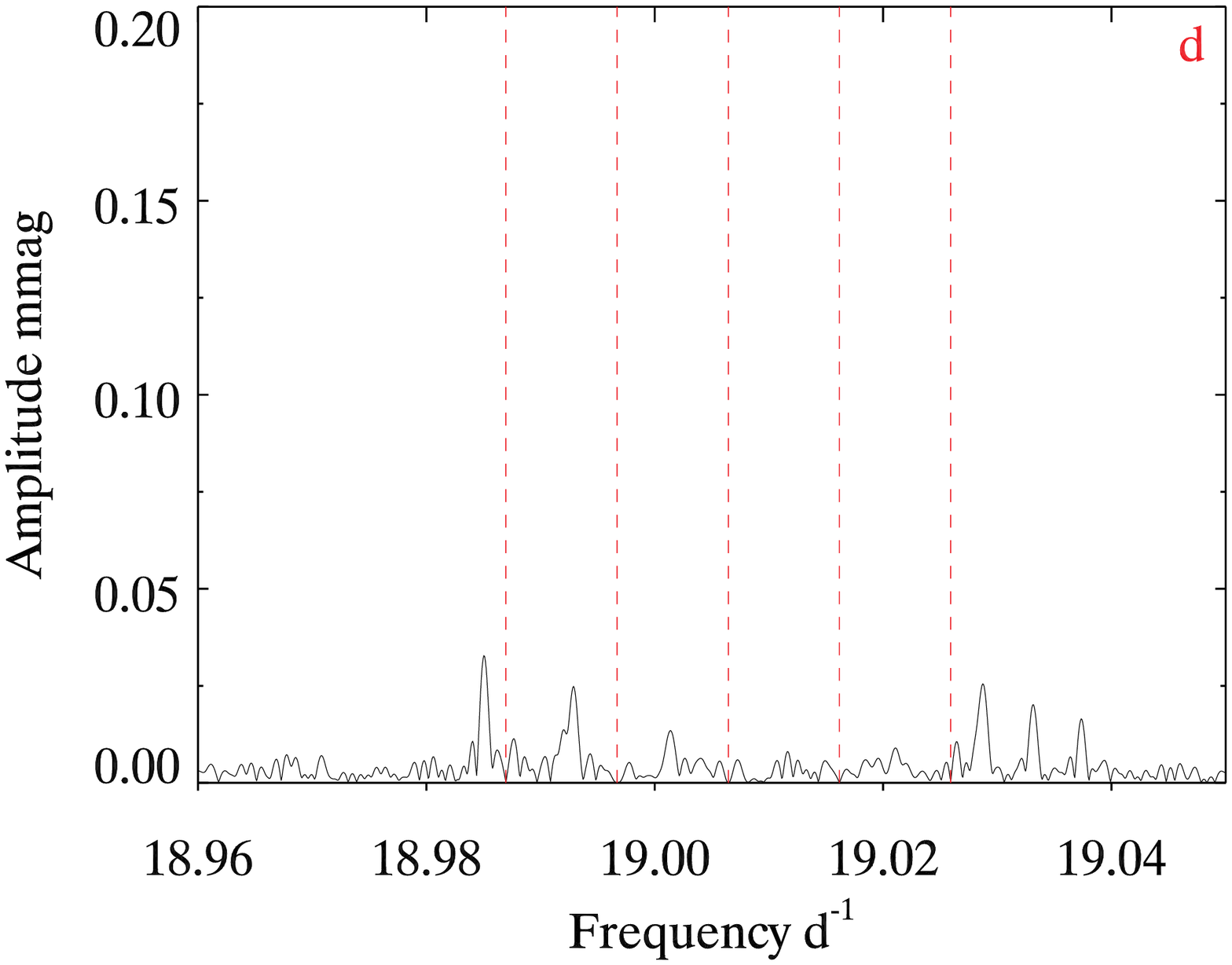}
\caption{(a): An amplitude spectrum for one of the p-mode triplets. (b): The amplitude spectrum of the residuals after prewhitening by the three triplet frequencies. (c): An amplitude spectrum for one of the p-mode quintuplets. (d): The amplitude spectrum of the residuals after prewhitening by the five quintuplet frequencies. Note the change of scale for the bottom panels.}
\label{fig:11145123_ft2}
\end{figure*}

Fig.\,\ref{fig:11145123_ft2} shows an example of a p-mode triplet and a p-mode quadrupole quintuplet. The low visibility of the central $m = 0$ mode peak for a dipole mode suggests a high inclination of the pulsation axis (which is assumed to be coincident with the rotation axis). The visibility of the $m = +1, -1$ modes of the quadrupole then argues that the inclination angle may be $i \sim 70^\circ$. This argument is not strong, since we do not expect equipartition of energy between modes in $\delta$\,Sct stars, but it is indicative of the orientation of the rotation axis to the line-of-sight. We will see in the next section that all of the g-mode dipole triplets show low visibility of the $m = 0$ mode. Some of the other p-mode triplets do not show this amplitude pattern.

\subsection{The g\,modes}
\label{sec:gmodes}

\begin{figure}
\centering
\includegraphics[width=0.9\linewidth,angle=0]{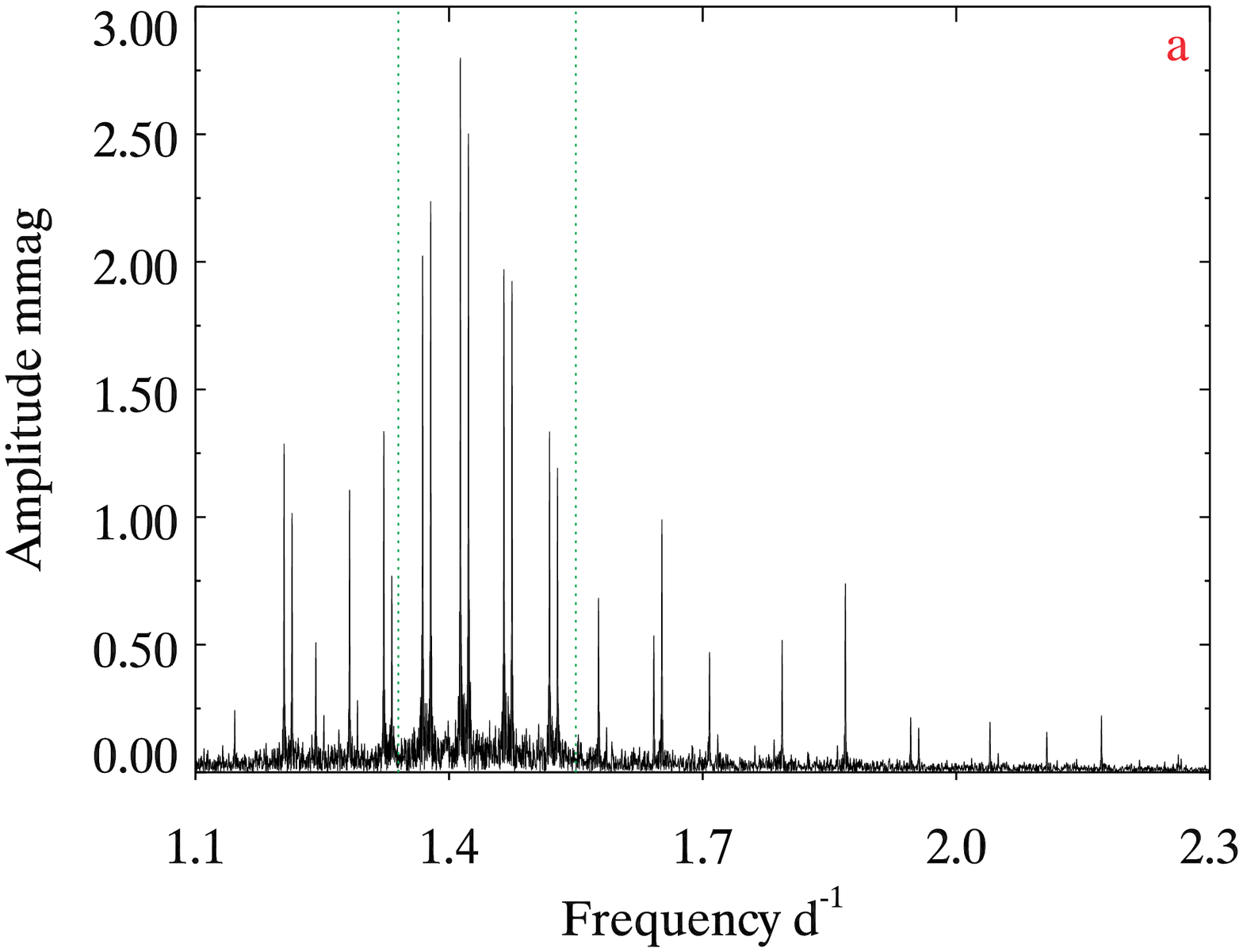}
\includegraphics[width=0.9\linewidth,angle=0]{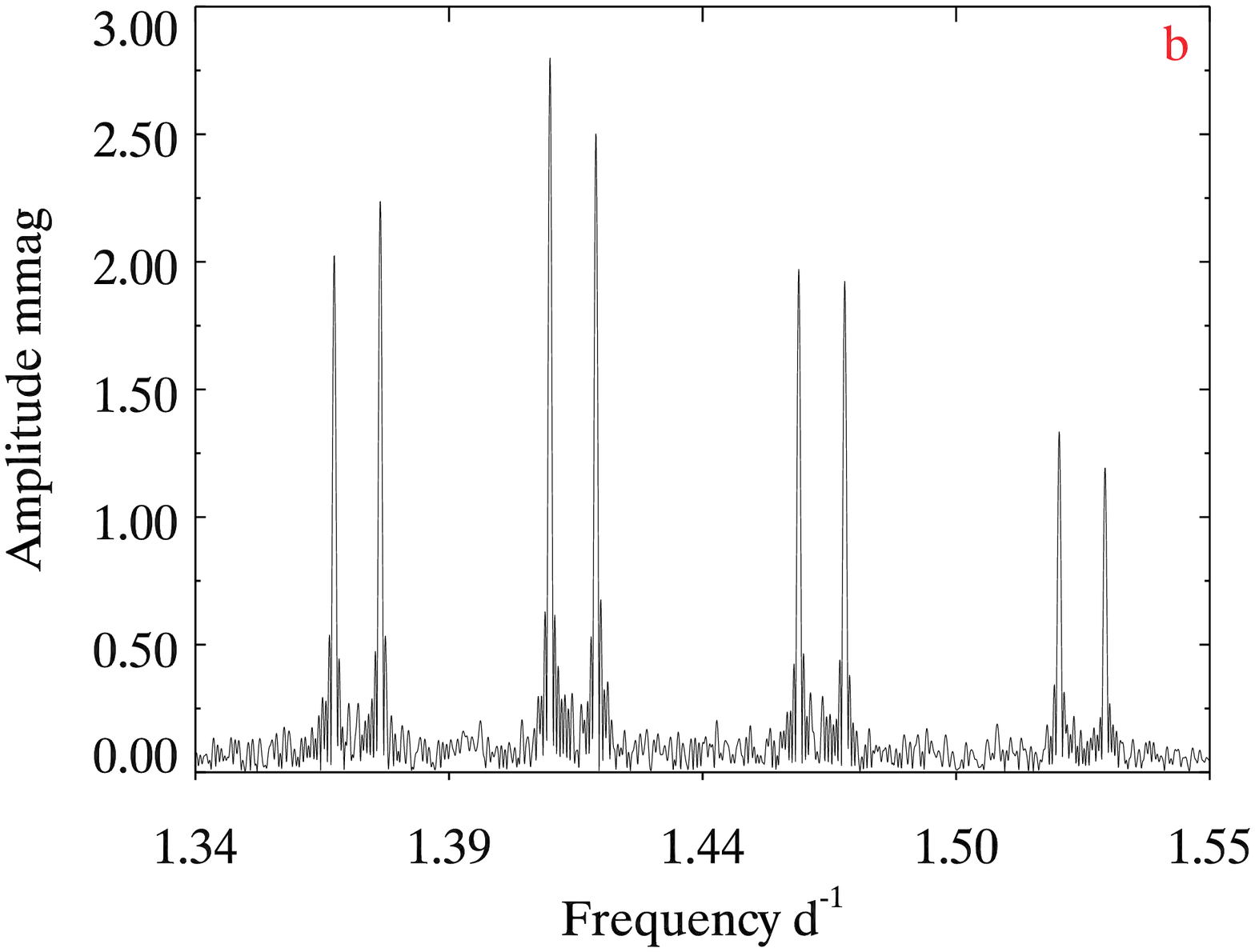}
\includegraphics[width=0.9\linewidth,angle=0]{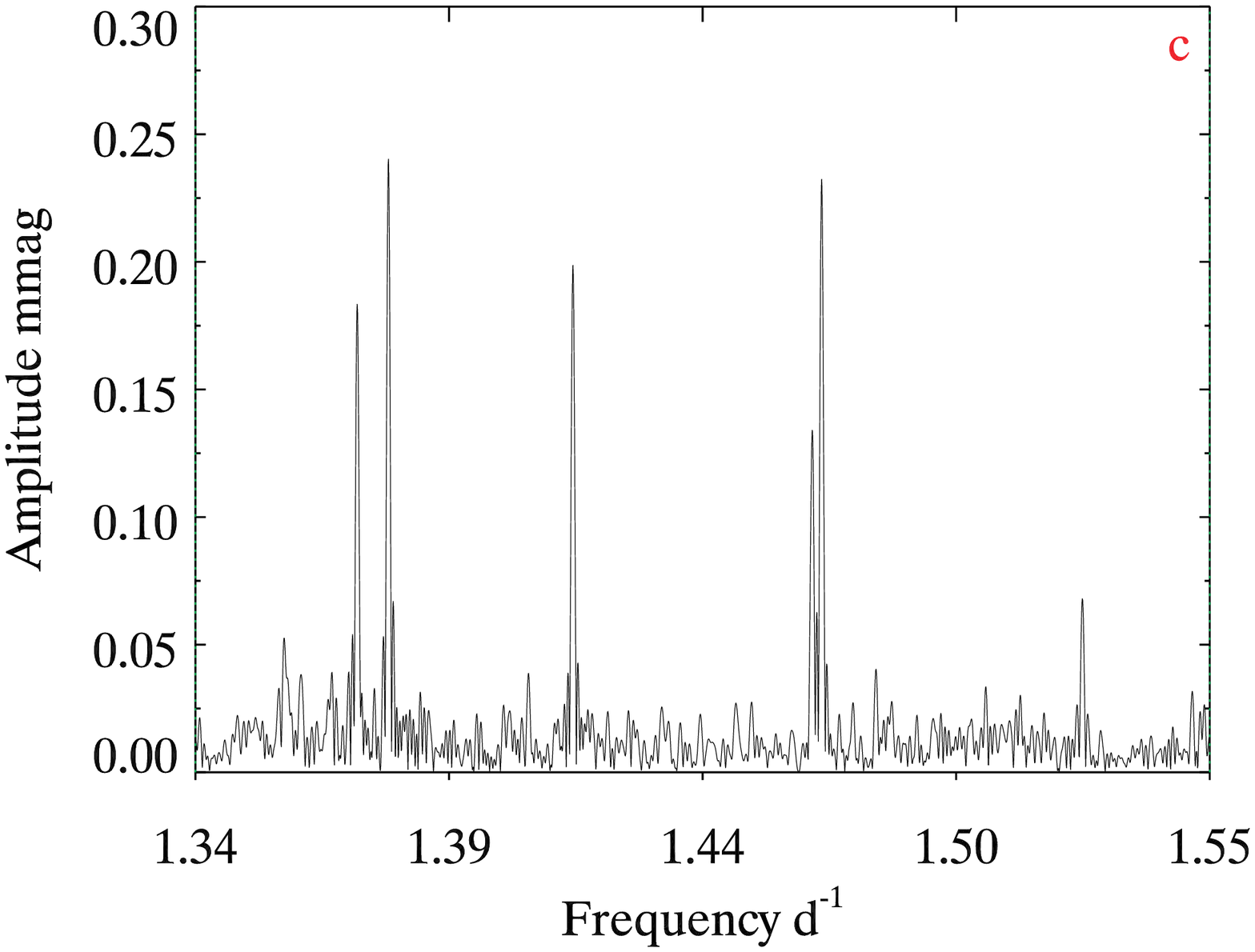}
\caption{(a): An amplitude spectrum for the g\,modes. There is a series of high overtone, nearly equally spaced mode triplets that appear at first look to be doublets. (b): A higher resolution look at the highest peaks. Panel (c) shows that the central peaks of the triplets are present; note the change of scale from the upper panels. }
\label{fig:11145123_ft3}
\end{figure}

Fig.\,\ref{fig:11145123_ft3} shows the g\,modes. A detailed examination of the amplitude spectrum shows at least 15 g-mode triplets, which we identify as consecutive overtones of dipole modes. All of the central $m = 0$ modes have low visibility. The higher resolution look at the four highest amplitude g-mode triplets shows that the $m=0$ mode peaks are visible. This again suggests that the inclination of the pulsation axis is high, as for the p-mode triplet shown in the previous section.

\subsection{The frequencies}
\label{sec:frequencies}

Tables\,\ref{table:11145123_g} and \ref{table:11145123_p} give the results of a combination of linear least-squares and nonlinear least-squares fits of 61 derived p-mode and g-mode frequencies to the Q0-16 data. Because of the decreasing signal to noise ratio for these further peaks, we chose for this first study of KIC\,11145123 to analyse only the most significant multiplets. There is still variance in the data, as we will see in section\,\ref{sec:modecouple} below, so the formal errors are slightly overestimated. They may, therefore, be considered to be conservative.

The first column in Tables\,\ref{table:11145123_g} and \ref{table:11145123_p} marks the g\,modes (g), and the p-mode singlet (s), triplets (t) and quintuplets (q). The next three columns give frequency, amplitude and phase (with respect to $t_0 = {\rm BJD}2455634.3$). The fifth column gives the frequency separation between the components of each multiplet. For the g\,modes half of the frequency splitting has been given, since the measured doublets are for the dipole $m = +1, -1$ modes, with the $m = 0$ mode either not visible, or not recorded here because of the low amplitude. For the p\,modes within the errors all splittings for a multiplet are equal, which is consistent with our expectation that there are no second-order effects for such a long rotation period ($\sim$100\,d), assuming that latitudinal differential rotation is insignificant.

While we model the internal rotation of KIC\,11145123 in section\,\ref{sec:internal_rotation} below, we note here that the ratio of the average of all of the p-mode splittings in Table\,\ref{table:11145123_p} to the average of all of the g-mode splittings in Table\,\ref{table:11145123_g} is 1.98. If the star rotates rigidly, a first-order value of 2 is expected for this ratio because the high-overtone g-mode Ledoux constant [cf.~equation\,(\ref{eq:cnl})] asymptotically approaches $C_{n,l} =  0.5$ (appendix \ref{appendix:Cn1_low_freq_lim}), and the p-mode Ledoux constants are expected to be close to zero. This already suggests almost uniform rotation of the star without detailed modelling of its structure. Thus we can determine the rotation in the deep interior and at the surface in a nearly model-independent manner. Our internal rotational measurements are {\it not model sensitive}.

\begin{table*}
\centering
\caption[]{A least squares fit of all of the frequency multiplets for KIC\,11145123. A total of 61 frequencies were fitted by nonlinear least squares. The time zero point is $t_0 = {\rm BJD} 2455634.3$. For clarity this table is split into two parts, although the fit was done simultaneously. This Table shows the g-mode frequencies.  }
\begin{tabular}{cccrc}
\hline
\hline
&\multicolumn{1}{c}{frequency} & \multicolumn{1}{c}{amplitude} &
\multicolumn{1}{c}{phase} & \multicolumn{1}{c}{$\Delta f$} \\
&\multicolumn{1}{c}{d$^{-1}$} & \multicolumn{1}{c}{mmag} &
\multicolumn{1}{c}{radians} & \multicolumn{1}{c}{d$^{-1}$} \\
\hline
g & $1.2048302 \pm 0.0000030$ & $1.3197 \pm 0.0095$ & $0.4888 \pm 0.0072$ &    \\
 g & $1.2143215 \pm 0.0000041$ & $0.9647 \pm 0.0095$ & $2.3206 \pm 0.0098$ & $0.0047456 \pm 0.0000025$ \\
\hline
g & $1.2423952 \pm 0.0000075$ & $0.5225 \pm 0.0095$ & $0.4202 \pm 0.0182$ &     \\
 g & $1.2519753 \pm 0.0000190$ & $0.2060 \pm 0.0095$ & $-1.6226 \pm 0.0460$ & $0.0047901 \pm 0.0000102$ \\
 \hline
g & $1.2823111 \pm 0.0000033$ & $1.1814 \pm 0.0095$ & $0.5905 \pm 0.0080$ &     \\
 g & $1.2918109 \pm 0.0000130$ & $0.3018 \pm 0.0095$ & $1.6393 \pm 0.0314$ & $0.0047499 \pm 0.0000067$ \\
\hline
 g & $1.3228337 \pm 0.0000031$ & $1.2679 \pm 0.0095$ & $0.7935 \pm 0.0075$  &   \\
 g & $1.3323722 \pm 0.0000056$ & $0.6970 \pm 0.0095$ & $1.9798 \pm 0.0136$ & $0.0047692 \pm 0.0000032$ \\
\hline
 g & $1.3687453 \pm 0.0000019$ & $2.0210 \pm 0.0095$ & $-0.4447 \pm 0.0047$ &    \\
 g & $1.3782838 \pm 0.0000017$ & $2.2364 \pm 0.0095$ & $-1.2430 \pm 0.0042$ & $0.0047693 \pm 0.0000013$ \\
\hline
 g & $1.4134084 \pm 0.0000014$ & $2.8147 \pm 0.0095$ & $0.4081 \pm 0.0034$ &    \\
 g & $1.4229207 \pm 0.0000016$ & $2.4954 \pm 0.0095$ & $-0.0643 \pm 0.0038$ & $0.0047562 \pm 0.0000010$ \\
\hline
 g & $1.4648931 \pm 0.0000019$ & $2.0234 \pm 0.0095$ & $-2.0308 \pm 0.0047$ &    \\
 g & $1.4744549 \pm 0.0000021$ & $1.8995 \pm 0.0095$ & $-0.8012 \pm 0.0050$ & $0.0047809 \pm 0.0000014$ \\
\hline
 g & $1.5188455 \pm 0.0000029$ & $1.3352 \pm 0.0095$ & $2.7598 \pm 0.0071$ &     \\
 g & $1.5283515 \pm 0.0000032$ & $1.2077 \pm 0.0095$ & $2.1678 \pm 0.0078$ & $0.0047530 \pm 0.0000022$ \\
\hline
 g & $1.5767857 \pm 0.0000057$ & $0.6825 \pm 0.0095$ & $0.5361 \pm 0.0139$ &     \\
 g & $1.5863571 \pm 0.0000209$ & $0.1870 \pm 0.0095$ & $2.0174 \pm 0.0507$ & $0.0047857 \pm 0.0000108$ \\
\hline
 g & $1.6422380 \pm 0.0000072$ & $0.5416 \pm 0.0095$ & $-2.3584 \pm 0.0175$ &    \\
 g & $1.6518053 \pm 0.0000040$ & $0.9658 \pm 0.0095$ & $-0.9337 \pm 0.0098$ & $0.0047836 \pm 0.0000041$ \\
\hline
 g & $1.7081867 \pm 0.0000080$ & $0.4857 \pm 0.0095$ & $-2.5744 \pm 0.0195$ &    \\
 g & $1.7177950 \pm 0.0000286$ & $0.1365 \pm 0.0095$ & $1.4300 \pm 0.0694$ & $0.0048041 \pm 0.0000149$ \\
\hline
 g & $1.7844364 \pm 0.0000323$ & $0.1213 \pm 0.0095$ & $2.3874 \pm 0.0783$ &    \\
 g & $1.7940049 \pm 0.0000078$ & $0.4994 \pm 0.0095$ & $-1.4506 \pm 0.0190$ & $0.0047842 \pm 0.0000166$ \\
\hline
 g & $1.8592729 \pm 0.0000479$ & $0.0817 \pm 0.0095$ & $0.1605 \pm 0.1160$ &    \\
 g & $1.8687875 \pm 0.0000054$ & $0.7246 \pm 0.0095$ & $-0.5083 \pm 0.0131$ & $0.0047573 \pm 0.0000241$ \\
\hline
 g & $1.9460793 \pm 0.0000174$ & $0.2247 \pm 0.0095$ & $-1.2539 \pm 0.0422$ &    \\
 g & $1.9556277 \pm 0.0000256$ & $0.1529 \pm 0.0095$ & $0.8132 \pm 0.0620$ & $0.0047742 \pm 0.0000155$ \\
\hline
 g & $2.0399136 \pm 0.0000203$ & $0.1928 \pm 0.0095$ & $1.0005 \pm 0.0491$ &    \\
 g & $2.0494919 \pm 0.0000528$ & $0.0741 \pm 0.0095$ & $0.0842 \pm 0.1280$ & $0.0047891 \pm 0.0000283$ \\
\hline
\hline
\end{tabular}
\label{table:11145123_g}
\end{table*}

\begin{table*}
\centering
\caption[]{A least squares fit of all of the frequency multiplets for KIC\,11145123. A total of 61 frequencies were fitted by nonlinear least squares. The time zero point is $t_0 = {\rm BJD} 2455634.3$. For clarity this table is split into two parts, although the fit was done simultaneously. This Table shows the p-mode frequencies.  }
\begin{tabular}{cccrc}
\hline
&\multicolumn{1}{c}{frequency} & \multicolumn{1}{c}{amplitude} &
\multicolumn{1}{c}{phase} & \multicolumn{1}{c}{$\Delta f$} \\
&\multicolumn{1}{c}{d$^{-1}$} & \multicolumn{1}{c}{mmag} &
\multicolumn{1}{c}{radians} & \multicolumn{1}{c}{d$^{-1}$} \\
\hline
\hline
 q & $16.7258824 \pm 0.0000017$ & $2.3298 \pm 0.0095$ & $1.4903 \pm 0.0041$ &    \\
 q & $16.7339455 \pm 0.0000186$ & $0.2111 \pm 0.0095$ & $2.6371 \pm 0.0450$ & $0.0080632 \pm 0.0000186$ \\
 q & $16.7420104 \pm 0.0000075$ & $0.5266 \pm 0.0095$ & $-0.8098 \pm 0.0180$ & $0.0080649 \pm 0.0000200$ \\
 q & $16.7500755 \pm 0.0000110$ & $0.3561 \pm 0.0095$ & $-0.9646 \pm 0.0266$ & $0.0080651 \pm 0.0000133$ \\
 q & $16.7580083 \pm 0.0000504$ & $0.0776 \pm 0.0095$ & $-0.2290 \pm 0.1221$ & $0.0079328 \pm 0.0000516$ \\
\hline
 s & $17.9635133 \pm 0.0000005$ & $7.2526 \pm 0.0095$ & $1.3127 \pm 0.0013$ &    \\
\hline
 t & $18.3558305 \pm 0.0000029$ & $1.3355 \pm 0.0095$ & $1.7401 \pm 0.0071$ &     \\
 t & $18.3660001 \pm 0.0000135$ & $0.2899 \pm 0.0095$ & $1.5984 \pm 0.0327$ & $0.0101696 \pm 0.0000138$ \\
 t & $18.3761210 \pm 0.0000034$ & $1.1325 \pm 0.0095$ & $-2.9178 \pm 0.0084$ & $0.0101209 \pm 0.0000139$ \\
\hline
 q & $18.9869603 \pm 0.0000078$ & $0.5021 \pm 0.0095$ & $-1.1092 \pm 0.0189$ &     \\
 q & $18.9967001 \pm 0.0000060$ & $0.6477 \pm 0.0095$ & $-0.9156 \pm 0.0146$ & $0.0097398 \pm 0.0000099$ \\
 q & $19.0064482 \pm 0.0000056$ & $0.6933 \pm 0.0095$ & $-2.4825 \pm 0.0137$ & $0.0097481 \pm 0.0000083$ \\
 q & $19.0161736 \pm 0.0000101$ & $0.3875 \pm 0.0095$ & $-0.0080 \pm 0.0244$ & $0.0097254 \pm 0.0000116$ \\
 q & $19.0259102 \pm 0.0000086$ & $0.4525 \pm 0.0095$ & $-2.6214 \pm 0.0209$ & $0.0097366 \pm 0.0000133$ \\
\hline
 t & $21.9933315 \pm 0.0000064$ & $0.6117 \pm 0.0095$ & $3.1067 \pm 0.0155$ &    \\
 t & $22.0018915 \pm 0.0000286$ & $0.1367 \pm 0.0095$ & $-1.1618 \pm 0.0693$ & $0.0085600 \pm 0.0000293$ \\
 t & $22.0104220 \pm 0.0000064$ & $0.6109 \pm 0.0095$ & $3.0405 \pm 0.0155$ & $0.0085305 \pm 0.0000293$ \\
\hline
 t & $23.5061953 \pm 0.0000545$ & $0.0719 \pm 0.0095$ & $-0.1717 \pm 0.1319$ &    \\
 t & $23.5160925 \pm 0.0001851$ & $0.0212 \pm 0.0095$ & $0.1104 \pm 0.4474$ & $0.0098972 \pm 0.0001930$ \\
 t & $23.5258540 \pm 0.0000224$ & $0.1747 \pm 0.0095$ & $-2.6177 \pm 0.0542$ & $0.0097615 \pm 0.0001865$ \\
\hline
 q & $23.5455350 \pm 0.0000303$ & $0.1293 \pm 0.0095$ & $1.0403 \pm 0.0733$ &    \\
 q & $23.5553428 \pm 0.0000582$ & $0.0676 \pm 0.0095$ & $0.5831 \pm 0.1408$ & $0.0098079 \pm 0.0000656$ \\
 q & $23.5651835 \pm 0.0000480$ & $0.0815 \pm 0.0095$ & $2.8241 \pm 0.1163$ & $0.0098407 \pm 0.0000754$ \\
 q & $23.5749885 \pm 0.0000607$ & $0.0648 \pm 0.0095$ & $-2.5933 \pm 0.1468$ & $0.0098050 \pm 0.0000774$ \\
 q & $23.5847898 \pm 0.0000163$ & $0.2403 \pm 0.0095$ & $0.5719 \pm 0.0394$ & $0.0098013 \pm 0.0000629$ \\
\hline
 t & $23.8082903 \pm 0.0001093$ & $0.0362 \pm 0.0095$ & $-3.0537 \pm 0.2650$ &    \\
 t & $23.8185035 \pm 0.0000701$ & $0.0560 \pm 0.0095$ & $2.4199 \pm 0.1699$ & $0.0102132 \pm 0.0001298$ \\
 t & $23.8286824 \pm 0.0000580$ & $0.0679 \pm 0.0095$ & $1.3008 \pm 0.1409$ & $0.0101790 \pm 0.0000910$ \\
\hline
 t & $24.4093984 \pm 0.0000173$ & $0.2245 \pm 0.0095$ & $1.6409 \pm 0.0420$ &     \\
 t & $24.4192854 \pm 0.0000559$ & $0.0699 \pm 0.0095$ & $0.3261 \pm 0.1355$ & $0.0098870 \pm 0.0000585$ \\
 t & $24.4291655 \pm 0.0000551$ & $0.0710 \pm 0.0095$ & $-2.3962 \pm 0.1333$ & $0.0098801 \pm 0.0000785$ \\
 \hline
\hline
\end{tabular}
\label{table:11145123_p}
\end{table*}

Because the 15 consecutive g-mode multiplets we present in Table\,\ref{table:11145123_g} are high radial overtone, they asymptotically approach equal period spacing. We show that and calculate the average period spacing for our model in Table\,\ref{table:periods}.

\begin{table*}
\centering
\caption[]{The central g-mode frequencies for each more triplet calculated by averaging the two observed $m = +1, -1$ mode frequencies are given in column 1. The second column then gives the corresponding period, and the final column gives the period spacing with the next mode period. The average period spacing is $0.0241 \pm 0.0009$\,d.  }
\begin{tabular}{ccc}
\hline
\multicolumn{1}{c}{frequency} & \multicolumn{1}{c}{period (P)} &
\multicolumn{1}{c}{$\Delta P$} \\
\multicolumn{1}{c}{d$^{-1}$} & \multicolumn{1}{c}{d} &
\multicolumn{1}{c}{d} \\
\hline
\hline
$1.2095758 \pm 0.0000030$ & $0.8267361 \pm 0.0000021$ & $0.0249306 \pm 0.0000084$ \\
$1.2471853 \pm 0.0000127$ & $0.8018055 \pm 0.0000082$ & $0.0248415 \pm 0.0000097$ \\
$1.2870610 \pm 0.0000086$ & $0.7769640 \pm 0.0000052$ & $0.0237267 \pm 0.0000057$ \\
$1.3276030 \pm 0.0000042$ & $0.7532372 \pm 0.0000024$ & $0.0251780 \pm 0.0000026$ \\
$1.3735146 \pm 0.0000018$ & $0.7280593 \pm 0.0000009$ & $0.0229225 \pm 0.0000012$ \\
$1.4181646 \pm 0.0000015$ & $0.7051368 \pm 0.0000007$ & $0.0247138 \pm 0.0000012$ \\
$1.4696740 \pm 0.0000021$ & $0.6804230 \pm 0.0000010$ & $0.0240821 \pm 0.0000017$ \\
$1.5235985 \pm 0.0000033$ & $0.6563409 \pm 0.0000014$ & $0.0240583 \pm 0.0000070$ \\
$1.5815714 \pm 0.0000171$ & $0.6322825 \pm 0.0000069$ & $0.0251260 \pm 0.0000073$ \\
$1.6470217 \pm 0.0000068$ & $0.6071566 \pm 0.0000025$ & $0.0233823 \pm 0.0000090$ \\
$1.7129908 \pm 0.0000255$ & $0.5837743 \pm 0.0000087$ & $0.0248717 \pm 0.0000127$ \\
$1.7892207 \pm 0.0000297$ & $0.5589026 \pm 0.0000093$ & $0.0224306 \pm 0.0000159$ \\
$1.8640302 \pm 0.0000449$ & $0.5364720 \pm 0.0000129$ & $0.0238758 \pm 0.0000152$ \\
$1.9508535 \pm 0.0000302$ & $0.5125961 \pm 0.0000079$ & $0.0235275 \pm 0.0000159$ \\
$2.0447027 \pm 0.0000578$ & $0.4890687 \pm 0.0000138$ &  \\
\hline
\hline
\end{tabular}
\label{table:periods}
\end{table*}

\subsection{Mode coupling}
\label{sec:modecouple}

\begin{figure}
\centering
\includegraphics[width=0.9\linewidth,angle=0]{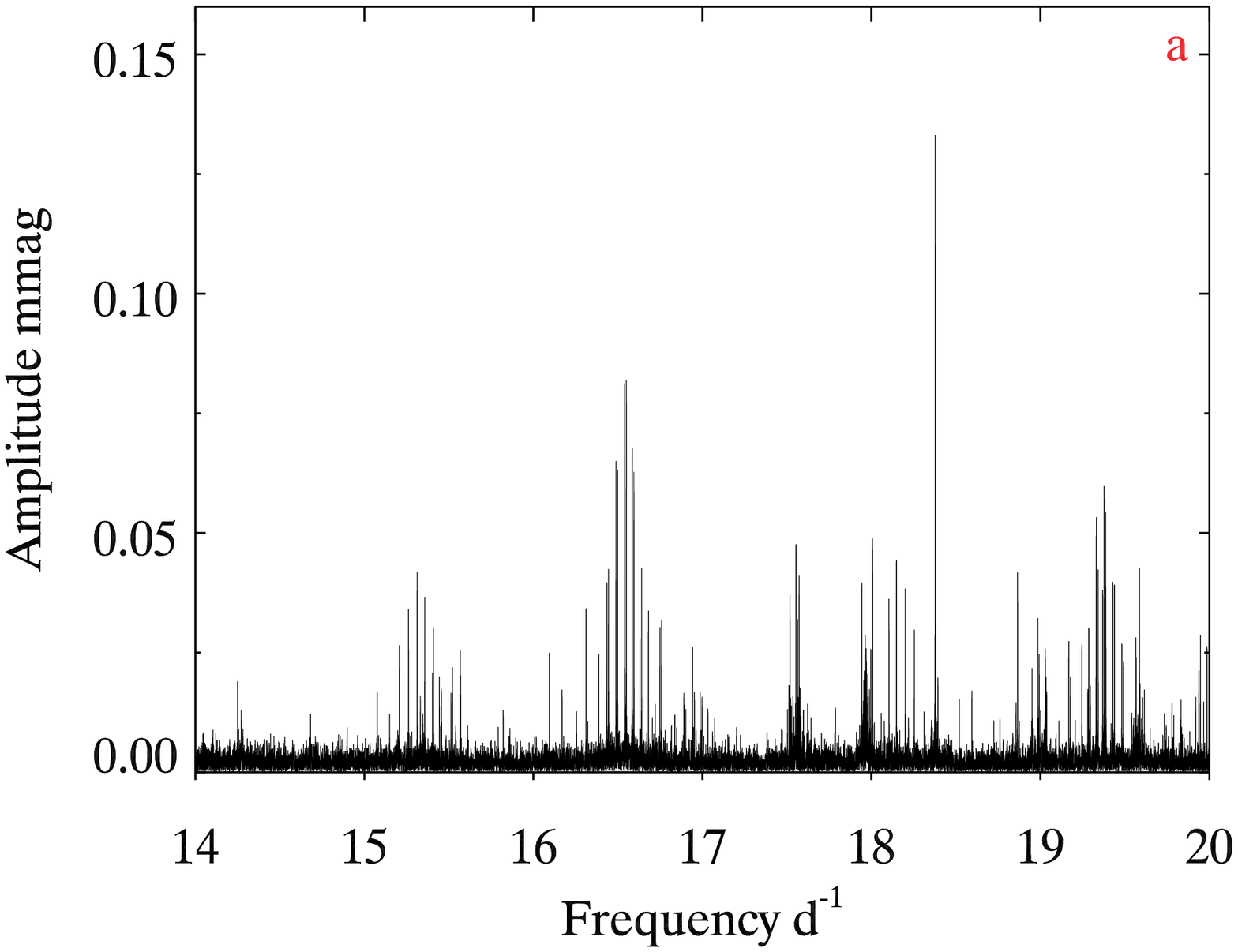}
\includegraphics[width=0.9\linewidth,angle=0]{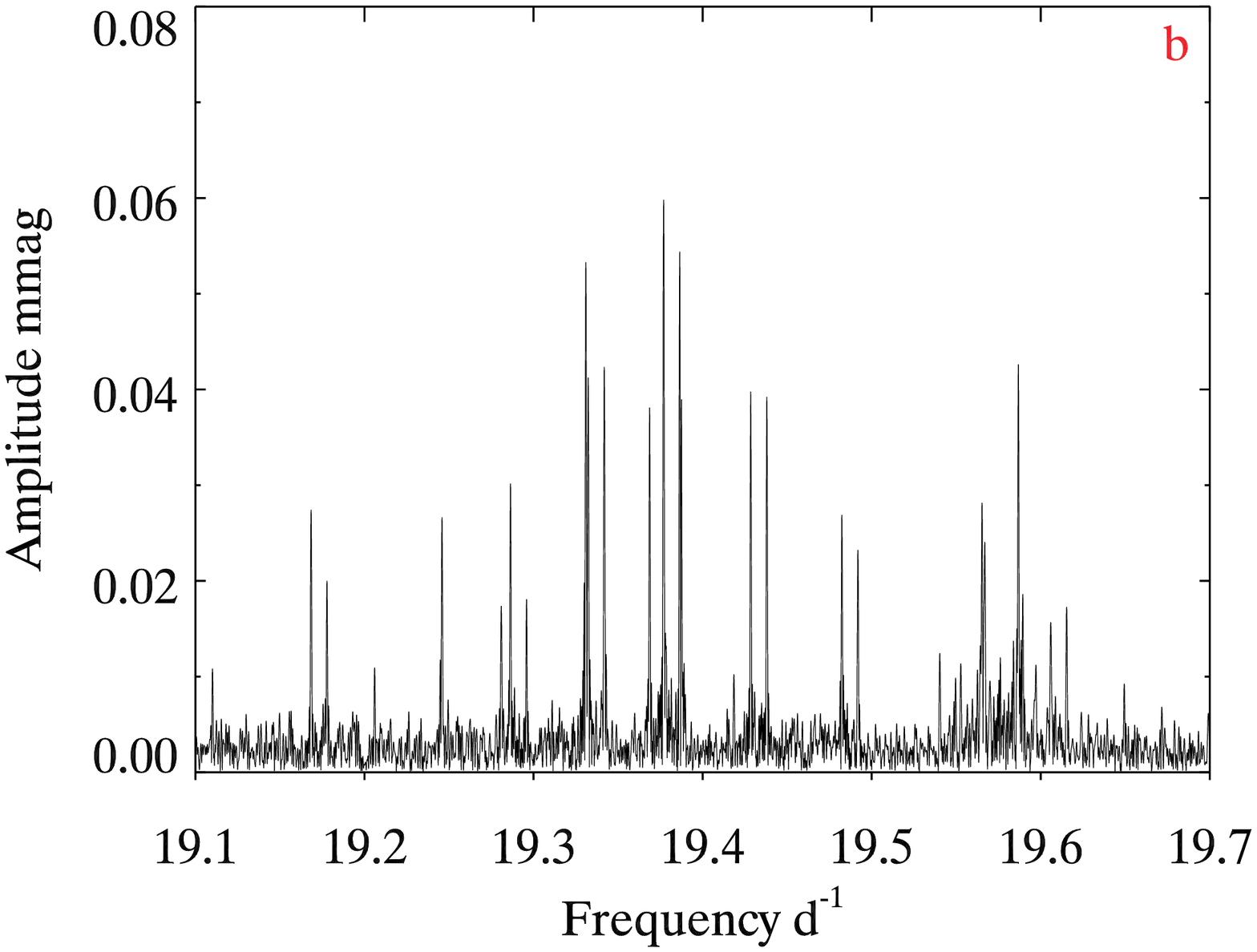}
\includegraphics[width=0.9\linewidth,angle=0]{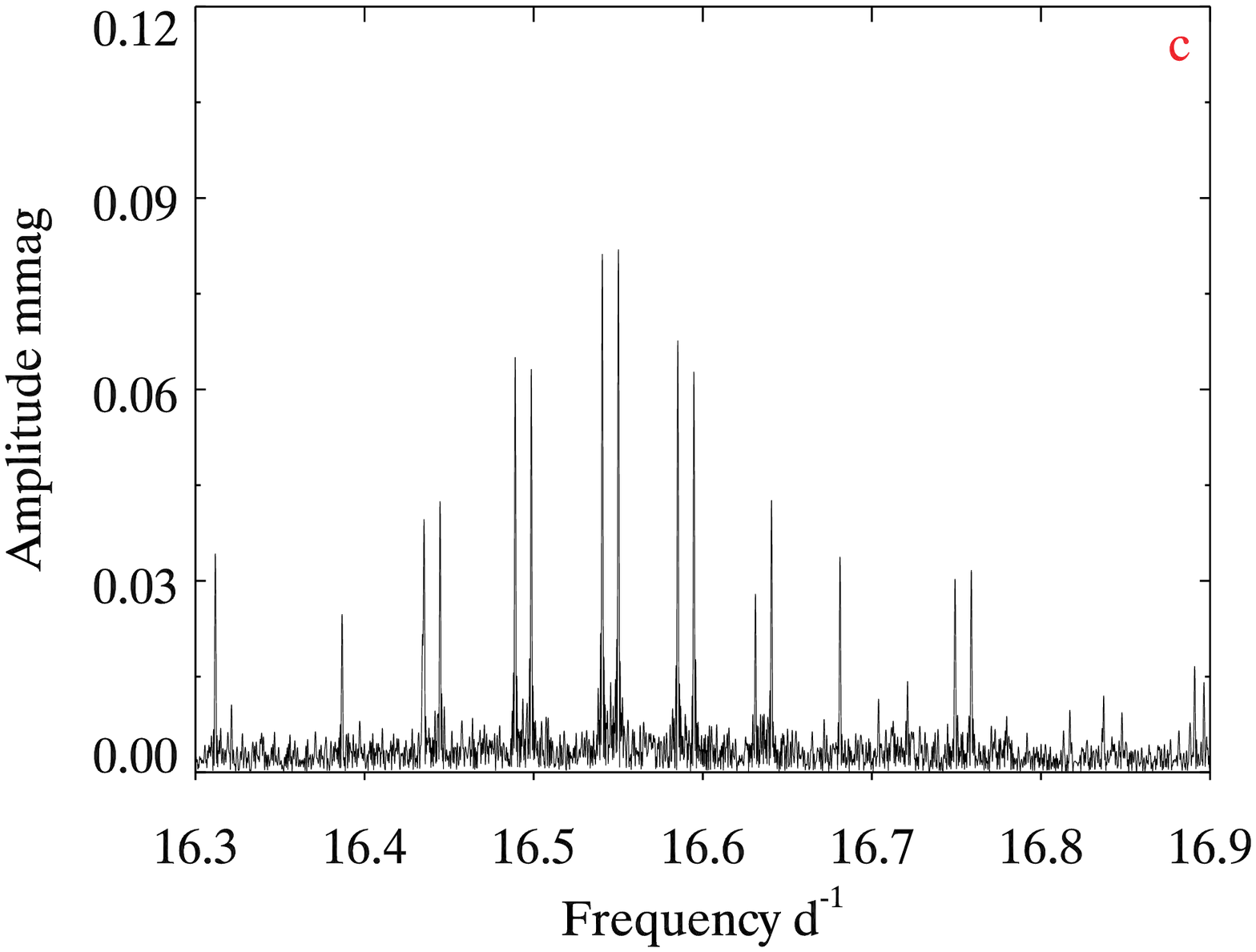}
\caption{(a): An amplitude spectrum for the p\,mode range after removing the peaks given in Tables\,\ref{table:11145123_g} and \ref{table:11145123_p}. Panels (b) and (c) show $\nu_1$ (the radial mode singlet highest amplitude mode) plus and minus the g\,mode pattern, respectively. This proves the p\,modes and g\,modes are coupled. }
\label{fig:11145123_ft4}
\end{figure}

Fig.\,\ref{fig:11145123_ft4} shows the combination frequencies,
which can be expressed by $\nu_1 \pm \nu_{\rm g}$ 
with $\nu_1$ and $\nu_{\rm g}$
being the frequency of the highest amplitude singlet p mode and 
that of each g mode, respectively.
These combination frequencies can
naturally be explained by nonlinear effects that occur when the p mode
and the g modes are excited simultaneously.  In fact, the fact that the
g-mode frequency distributions are precisely reproduced in both sides of
$\nu_1$ is fully consistent with this interpretation.  Therefore, the
presence of the combination frequencies proves that both p modes and g
modes originate in the same star.

\section{Model}
\label{sec:model}

Theoretical models of KIC\,11145123 are constructed in this section. We have calculated evolutionary models using the MESA (Modules for Experiments in Stellar Astrophysics; version 4298) code \citep{paxton2013}, and have performed linear adiabatic pulsation analyses for the models using pulsation codes based on those of \citet{saio1980} and \citet{takata2012}. The heavy element abundance is scaled by the solar mixture of \citet{asplund2009}, and OPAL opacity tables \citep{opal} are used. The mixing length is set equal to 1.7\,$H_p$ with $H_p$ being  the pressure scale height, although changing the mixing length hardly affects the results discussed in this paper. Atomic diffusion is activated in the code to erase noise in the distribution of the Brunt-V\"ais\"al\"a frequency in the zones with mean molecular weight gradients.  Although atomic diffusion causes the helium abundance, $Y$, in the outermost layers to decrease by about $0.08 - 0.15$ at the end of main-sequence stage, it has been checked that this hardly changes the property of low-order p\,modes which we compare with observation.

\subsection{Period spacing of high-order g\,modes}

One of the remarkable pulsation properties of KIC\,11145123 is the regular frequency spacing of triplet g\,modes, given in Table\,\ref{table:periods},  which is expected from the asymptotic theory of g\,modes. The average period spacing of the central frequencies of the 15 consecutive high-overtone g-mode  triplets is $\Delta P_{\rm g} = 0.024$\,d, as can be seen in Table\,\ref{table:periods}. This period spacing is asymptotically proportional to $[\int N d \ln r]^{-1}$, where $N$ is the Brunt-V\"ais\"al\"a (or buoyancy) frequency.

\begin{figure}
\centering
\includegraphics[width=0.95\linewidth,angle=0]{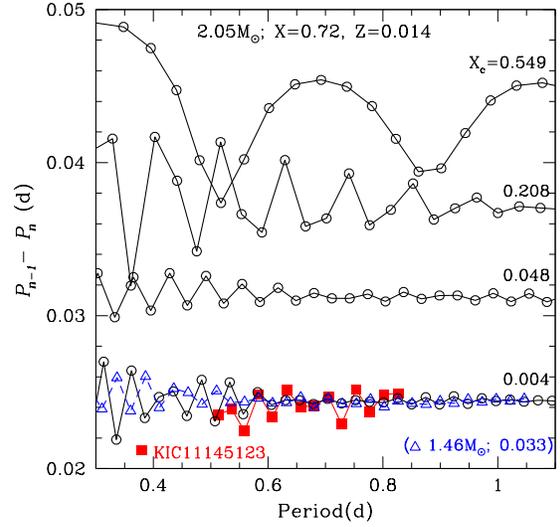}
\caption{Period spacings of g\,modes ($P_{n-1}-P_n$) as a function of g-mode period for selected evolutionary stages of the 2.05-M$_{\odot}$ model with $(X,Z)=(0.72,0.014)$. The observed period spacings of KIC\,11145123 are shown by filled squares. Also plotted (triangles) are period spacings of the best model of 1.46\,M$_{\odot}$ with $(X,Z)=(0.65,0.010)$. This shows KIC\,11145123 to be at the TAMS stage with core hydrogen heavily depleted.}
\label{fig:pdp}
\end{figure}

\begin{figure}
\centering
\includegraphics[width=0.9\linewidth,angle=0]{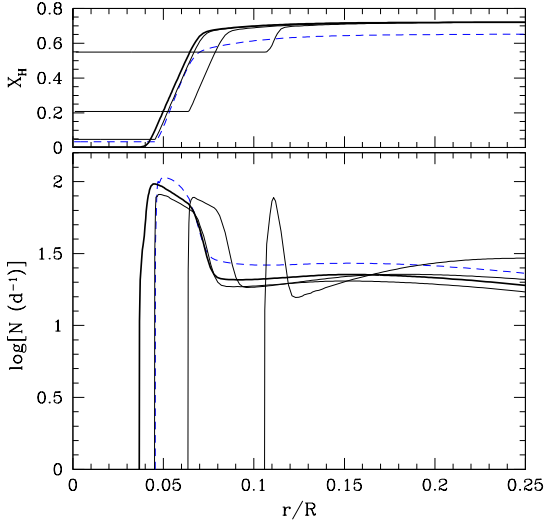}
\caption{Lower panel: Brunt-V\"ais\"al\"a frequency in units of d$^{-1}$ as a function of fractional radius at stages corresponding to those in Fig.\,\ref{fig:pdp}. Upper panel: Hydrogen mass fraction as a function of fractional radius; each line corresponds to the stage in the lower panel. Solid and dashed lines are for  2.05\,M$_{\odot}$ with $(X,Z)=(0.72,0.014)$ and for  1.46\,M$_{\odot}$ with $(X,Z)=(0.65,0.010)$, respectively.}
\label{fig:bvf_xh}
\end{figure}

As a star evolves, the distribution of the Brunt-V\"ais\"al\"a frequency in the interior varies, hence g-mode period spacings vary. Fig.\,\ref{fig:pdp} shows the period spacing $(P_{n-1}-P_n)$ as a function of g-mode period at selected evolutionary stages of  the 2.05-M$_{\odot}$ model with $(X,Z)=(0.72,0.014)$. As discussed by \citet{miglio2008}, the period spacing is modulated due to the presence of a steep variation in the Brunt-V\"ais\"al\"a frequency caused by a change in the gradient of the distribution of mean molecular weight. Fig.\,\ref{fig:bvf_xh} shows the distributions of hydrogen abundance and the Brunt-V\"ais\"al\"a frequency for each of the evolution stages shown in Fig.\,\ref{fig:pdp}. The modulation is significant in relatively early phases due to a sharp variation in the Brunt-V\"ais\"al\"a frequency. The modulation period and its amplitude, as well as the mean spacing $\Delta P_{\rm g}$, decrease as the evolution proceeds. The terminal age main sequence (TAMS) contraction starts when the central hydrogen abundance is reduced to $X_{\rm c}\approx 0.05$. The period spacings of KIC\,11145123, which hardly modulate, are consistent with a model in the TAMS contraction phase.

\begin{figure}
\centering
\includegraphics[width=0.9\linewidth,angle=0]{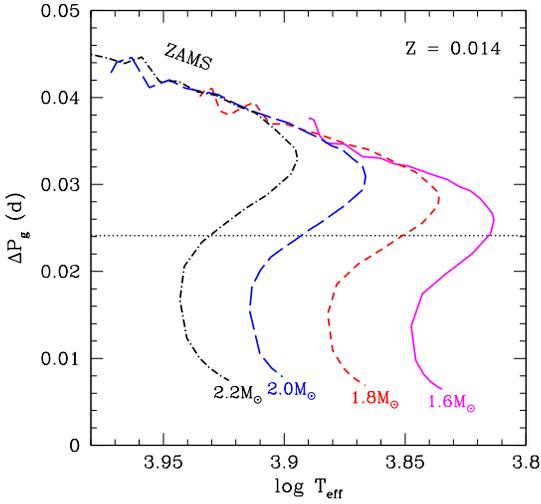}
\caption{Variations of mean period spacing, $\Delta P_{\rm g}$,  with evolution from the ZAMS to the end of the TAMS contraction for selected masses. At each evolutionary stage, $\Delta P_{\rm g}$ was obtained by averaging period the differences ($P_{n-1}-P_n$) in the observed period range of KIC\,11145123 ($0.48 - 0.83$\,d). The wiggles seen in relatively early phases are due to large modulation in the spacings as seen in Fig.\,\ref{fig:pdp}. The period spacing decreases as the central mass concentration increases with evolution. The TAMS contraction starts at minimum effective temperature and ends at a local maximum.  The g-mode period spacing of KIC\,11145123, 0.024\,d, is realized only during the TAMS contraction phase.}
\label{fig:gdp}
\end{figure}

Fig.\,\ref{fig:gdp} shows how the mean period spacing of g-modes varies during the evolution for various masses.  As the star evolves, the mean spacing $\Delta P_{\rm g}$ decreases, gradually at first, then rapidly after  the  TAMS  contraction phase. The value of $\Delta P_{\rm g}$ can therefore be regarded as an indicator of the evolutionary stage of the star. The period spacing of KIC\,11145123,  $\Delta P_{\rm g} = 0.024\,{\rm d}$, can be reproduced during  the TAMS contraction phase  in the mass range appropriate for $\delta$\,Sct -- $\gamma$\,Dor stars;  this is a robust constraint on the model of KIC\,11145123. The position where $\Delta P_{\rm g}=0.024$\,d in the HR diagram is indicated by a filled circle in Fig.\,\ref{fig:hrd} on each evolutionary track.

\subsection{Radial order of the singlet at $\nu_1$}

\begin{figure}
\centering
\includegraphics[width=0.9\linewidth,angle=0]{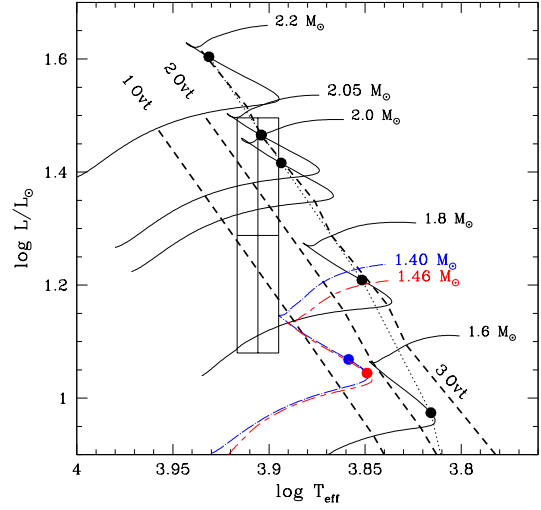}
\caption{An HR diagram for main sequence stars with masses consistent with the effective temperature and surface gravity of KIC\,11145123. The box shows the KIC estimates $T_{\rm  eff} = 8050 \pm 200$\,K and $\log g  = 4.0 \pm 0.2$ with 1$\sigma$ errors. The evolutionary tracks of solid lines are for a solar metal abundance of $Z = 0.014$, and an initial helium abundance of $Y = 0.266$. The filled circles are the positions where the average period spacing of the g\,modes is $\Delta P_{\rm g} = 0.024$\,d, as observed; they are connected by a dotted line for clarity. The dashed lines show the position of stars for which our models fit the highest amplitude singlet p\,mode seen in Fig.\,\ref{fig:11145123_ft-all} as a radial first, second or third overtone mode.  The $M = 2.05$-M$_{\odot}$ model of the standard abundance fits the singlet as the radial third-overtone, and the g-mode period spacing, as well as the KIC $T_{\rm  eff}$ and $\log g$, although the model fails to fit some p\,mode frequencies.  The evolutionary tracks with a dash-dotted line and with a long-dash-short-dash line are  for 1.40\,M$_{\odot}$ with $(X,Z)=(0.58,0.014)$ and for 1.46\,M$_{\odot}$ with $(X,Z)=(0.65,0.010)$, respectively. Models at filled circles on these evolutionary tracks are best models in the sense that they reproduce reasonably well all observed p-mode frequencies as well as $\Delta P_{\rm g}$. }
\label{fig:hrd}
\end{figure}

We identify the highest-amplitude p-mode singlet at $17.964\,{\rm d}^{-1}$ ($=\nu_1$), which is shown in Fig.\,\ref{fig:11145123_ft-all} and Table\,\ref{table:11145123_p}, as a radial mode. This is another important constraint on the model of KIC\,11145123. Dashed lines in Fig.\,\ref{fig:hrd} indicate approximate loci  where radial first, second, and third overtones have the frequency $\nu_1$. If a filled circle on an evolutionary track is located on a dashed line, the model at the filled circle reproduces $\Delta P_{\rm g}$ as well as the singlet frequency $\nu_1$.  Fig.\,\ref{fig:hrd} shows that for models of  $2.0 \la M/{\rm M}_\odot \la 2.2$ with  our standard initial composition $(X,Z)=(0.72,0.014)$, the points where $\Delta P_{\rm g}=0.024$\,d are on the dashed line  for the third-overtone radial-mode. Among the models, the model of 2.05\,M$_{\odot}$  lies well within the error box based on KIC parameters revised by \citet{huber2014}, i.e., $T_{\rm  eff} = 8050 \pm 200$\,K and $\log g = 4.0 \pm 0.2$.

If a different initial chemical composition is adopted, and/or  a small overshooting from a convective-core boundary is taken into account, we can obtain models which fit (in addition to the observed $\Delta P_{\rm g}$) the singlet frequency $\nu_1$ with a second- or a fourth-overtone radial-mode. For example, the singlet frequency $\nu_1$ is reproduced by the second-overtone radial-mode of the model,  $(M/{\rm M}_\odot, X,Z) = (1.40, 0.58, 0.014)$ or  $(1.46,0.65,0.010)$, with $\Delta P_{\rm g} = 0.024$\,d  as seen in Fig.\,\ref{fig:hrd}. (In this figure, the filled circles for these models are not on the dashed line for the second overtone because of the composition differences.) In summary, there are many models which reproduce $\Delta P_{\rm g}$ and $\nu_1$. From them we choose a few best models by using frequencies of nonradial p-modes of KIC\,11145123.

\subsection{Fitting with other p\,modes}

\begin{figure}
\centering
\includegraphics[width=0.9\linewidth,angle=0]{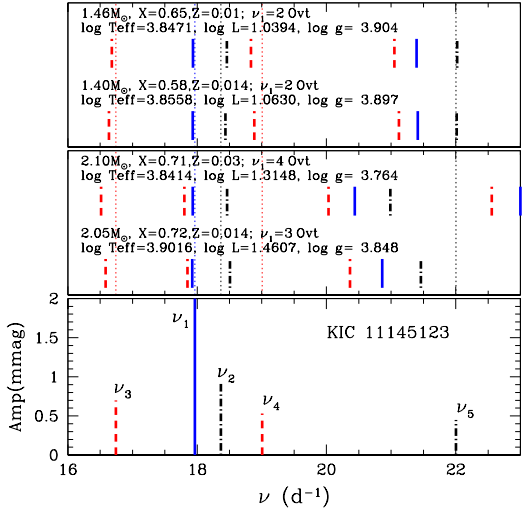}
\caption{Observed p-mode frequencies (bottom panel) are compared with frequencies of selected models (top and middle panels). Solid lines, dash-dotted lines, and dashed lines are for radial modes, dipole ($l=1$) modes, and quadrupole ($l=2$) modes, respectively.  }
\label{fig:pmodes}
\end{figure}

Fig.\,\ref{fig:pmodes} compares observed p-mode frequencies (bottom panel) with  corresponding frequencies from selected models (relatively massive models in the middle panel, and less massive models in the top panel). The 2.05-M$_{\odot}$ model in the middle panel has the standard initial composition, and lies within the error box on the HR diagram in Fig.\,\ref{fig:hrd}. Although this model reproduces reasonably well
$\nu_1$ (radial third-overtone),
$\nu_2$ ($l=1$, $18.366\,{\rm d}^{-1}$) and
$\nu_3$ ($l=2$, $16.742\,{\rm d}^{-1}$),  it cannot reproduce
$\nu_4$ ($l=2$, $19.006\,{\rm d}^{-1}$) and
$\nu_5$ ($l=1$, $22.002\,{\rm d}^{-1}$).
The situation is similar for the metal-rich ($Z=0.03$) 2.10-M$_{\odot}$  model in the middle panel, except that $\nu_1$ corresponds to the radial fourth-overtone in this case. In particular, it is remarkable that the frequency of the $l=2$ mode thought to correspond to $\nu_4 ( > \nu_2)$ is lower than $\nu_2$ in these models.

To obtain models which fit p-mode frequencies better than the above models, we have calculated various other models that reproduce $\Delta P_{\rm g}$ and $\nu_1$. A model with a given mass and an initial chemical composition is found which has $\Delta P_{\rm g}$ that agrees with the observed value ($0.024\,{\rm d}$) during the TAMS contraction phase. At that stage, however, the model does not necessarily have a radial-mode period consistent with $\nu_1$. By repeating above calculations with different initial helium abundances, we obtain a model that reproduces both $\Delta P_{\rm g}$ and $\nu_1$ for a given mass and metallicity $Z$.

\begin{figure}
\centering
\includegraphics[width=0.9\linewidth,angle=0]{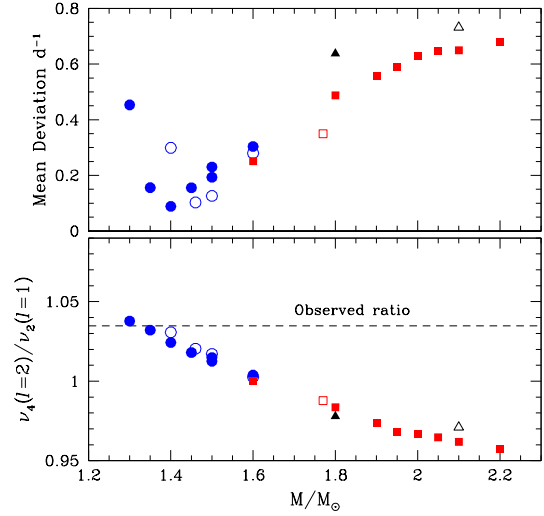}
\caption{Mean deviation of the four nonradial modes (upper panel) and predicted ratio for $\nu_4/\nu_2$ (lower panel) as functions of stellar mass. All models plotted are in the TAMS contraction phase, reproducing the g-mode period spacing, $\Delta P_{\rm g}$, as well as the singlet frequency, $\nu_1$. The singlet is identified as the second overtone (circles), the third overtone (squares), or the fourth overtone (triangles) depending on the models. Filled symbols stand for models with $Z=0.014$, open circles for models with $Z=0.010$, open squares for models with $Z=0.020$, open triangles for models with $Z=0.03$. The ratio $\nu_4/\nu_2$ is calculated by $\omega(n,l=2)/\omega(n+2,l=1)$ with the radial order $n$  set to $n_{\rm s} - 3$. Here, $n_{\rm s}$ is the radial order of the singlet $\nu_1$. (Note that the radial order of the $k$-th radial-overtone is equal to $k+1$.)}
\label{fig:ratio}
\end{figure}

The nonradial p-mode frequencies of the models thus obtained are compared with the observed ones; the frequency ratio corresponding to $\nu_4/\nu_2$ and the mean deviation from the observed four nonradial p-modes are plotted in Fig.\,\ref{fig:ratio} as functions of the model mass. Different symbols distinguish the radial overtone fitted to $\nu_1$ and the metallicity of the models. Generally, the singlet $\nu_1$ is fitted to the second overtone in less massive models, while it is fitted to the third or fourth overtone in massive models. This figure shows that the $\nu_4/\nu_2$ ratio (lower panel) and the mean deviation depend almost entirely on the model mass, with little dependence on metallicity and the radial order fitted to $\nu_1$. At $\sim$2\,M$_{\odot}$ the $\nu_4/\nu_2$ ratio is much smaller than the observed ratio. As the mass decreases the frequency ratio approaches the observed $\nu_4/\nu_2$ ratio and the mean deviation decreases; this indicates that in order to fit all of the observed p\,modes, the stellar mass must be considerably smaller than 2\,M$_{\odot}$ despite the KIC parameters for the position in the HR diagram.

The mean deviation attains a minimum at 1.40\,M$_{\odot}$  for $Z=0.014$ and 1.46\,M$_{\odot}$ for $Z=0.010$; they have initial helium abundances of  0.406 and 0.36, respectively, considerably higher than the normal abundance. Fig.\,\ref{fig:pmodes} shows that p-mode frequencies of these models (top panel) reproduce reasonably well all the observed p-mode frequencies with large amplitude. They are our best models of KIC\,11145123 based on the p- and g-mode frequencies. Table \ref{table:mode_id} provides mode identification based on the best model with 1.46\,M$_{\odot}$. While all of the g\,modes and most of the p\,modes are successfully identified, two of the low-amplitude p\,modes (those with frequencies $23.516\,{\rm d}^{-1}$ and $23.819\,{\rm d}^{-1}$), which are not taken into account in the best-model search, cannot be identified with modes with $l\le 2$. We tentatively identify them as those with $l=6$, which is usually too large to be detected because of large geometrical cancellation.

Although our best models reasonably agree with the observed frequencies, we note that the luminosities and effective temperatures of these models are outside (too faint and too cool) of the error box as seen in Fig.\,\ref{fig:hrd}. Given that in our models high helium abundance is required in the envelope, KIC\,11145123 could be an SX\,Phe variable that was formed by a close encounter of two stars in a dense stellar cluster. To test this idea a detailed spectroscopic study is needed to measure surface CNO abundances, as well as other heavy element abundances of KIC\,11145123.

\begin{table}
\centering
\caption[]{Mode identification of the observed modes based on  our best model with 1.46\,M$_{\odot}$. The first column indicates the type of mode that is common to Tables \ref{table:11145123_g} and \ref{table:11145123_p}, while the fourth and fifth columns represent the spherical degree, $l$, and  the radial order, $n$, of each mode, respectively. The identification of two modes that have a question mark in the sixth column should be regarded as tentative. }
\begin{tabular}{ccccrcc}
\hline
&
\multicolumn{1}{c}{observed frequency} &
\multicolumn{1}{c}{model frequency} &
\multicolumn{1}{c}{$l$} &
\multicolumn{1}{c}{$n$} \\
&
\multicolumn{1}{c}{d$^{-1}$} &
\multicolumn{1}{c}{d$^{-1}$}
\\
\hline
\hline
g & $1.210$ & $1.209$ & $1$ & $-33$
\\
g & $1.247$ & $1.245$ & $1$ & $-32$
\\
g & $1.287$ & $1.284$ & $1$ & $-31$
\\
g & $1.328$ & $1.325$ & $1$ & $-30$
\\
g & $1.374$ & $1.369$ & $1$ & $-29$
\\
g & $1.418$ & $1.416$ & $1$ & $-28$
\\
g & $1.470$ & $1.467$ & $1$ & $-27$
\\
g & $1.524$ & $1.521$ & $1$ & $-26$
\\
g & $1.582$ & $1.579$ & $1$ & $-25$
\\
g & $1.647$ & $1.642$ & $1$ & $-24$
\\
g & $1.713$ & $1.712$ & $1$ & $-23$
\\
g & $1.789$ & $1.786$ & $1$ & $-22$
\\
g & $1.864$ & $1.868$ & $1$ & $-21$
\\
g & $1.951$ & $1.957$ & $1$ & $-20$
\\
g & $2.045$ & $2.057$ & $1$ & $-19$
\\
\hline
q ($\nu_3$) & $16.742$ & $16.693$ & $2$ & $-1$
\\
s ($\nu_1$) & $17.964$ & $17.925$ & $0$ & $3$
\\
t ($\nu_2$) & $18.366$ & $18.448$ & $1$ & $2$
\\
q ($\nu_4$) & $19.006$ & $18.849$ & $2$ & $0$
\\
t ($\nu_5$) & $22.002$ & $22.007$ & $1$ & $3$
\\
t & $23.516$ & $23.444$ & $6$ & $-2$ & ?
\\
q & $23.565$ & $23.488$ & $2$ & $2$
\\
t & $23.819$ & $23.966$ & $6$ & $-1$ & ?
\\
t & $24.419$ & $24.453$ & $2$ & $3$
\\
\hline
\hline
\end{tabular}
\label{table:mode_id}
\end{table}

\section{Internal rotation}
\label{sec:internal_rotation}

The internal rotation of KIC\,11145123 is analysed based on the observed frequency splittings. After providing basic formulae about the rotational splittings, our analysis proceeds in two steps that give {\it nearly model-independent inferences}, and two-zone modelling of the rotation profile.

\subsection{Rotational splittings}

If a star rotates slowly, the Coriolis force and advection perturb the eigenfrequencies of  its oscillations, whereas the effect of the deformation of the star caused by the centrifugal force is negligible. If the angular velocity, $\Omega$, depends only on the radius, $r$, then the frequency perturbation, $\delta\omega$, in the inertial frame of reference is given by
\begin{equation}
\delta\omega_{n,l,m} = m(1-C_{n,l}) \int_0^R K_{n,l}(r)\Omega(r) dr,
\label{eq:split}
\end{equation}
where $n,l,m$ are  the radial order, the spherical degree and the azimuthal order; $C_{n,l}$, which is  sometimes called the Ledoux constant \citep{ledoux1951}, is defined as
\begin{equation}
C_{n,l}=\frac{
\int_0^R\xi_{\rm h}(2\xi_r+\xi_{\rm h})r^2\rho dr}
{
\int_0^R\left[\xi_r^2+l(l+1)\xi_{\rm h}^2\right]r^2\rho dr},
\label{eq:cnl}
\end{equation}
with radial ($\xi_r$) and horizontal ($\xi_{\rm h}$) displacements;
$\rho$ and $r$ are the gas density and the distance from the centre,
respectively.
The kernel, $K_{n,l}$,  is given as
\begin{equation}
K_{n,l}=
\frac{\left[\xi_r^2+l(l+1)\xi_{\rm h}^2 - 2\xi_r\xi_{\rm h}-\xi_{\rm h}^2\right]\rho r^2}%
{
\int_0^R
\left[\xi_r^2+l(l+1)\xi_{\rm h}^2 - 2\xi_r\xi_{\rm h}-\xi_{\rm h}^2\right]\rho r^2 dr}.
\label{eq:kernel}
\end{equation}
Since
\begin{equation}
\int_0^R K_{n,l}(r)dr = 1,
\label{eq:4}
\end{equation}
the integral on the right hand side of equation\,(\ref{eq:split})  gives the rotation rate averaged with respect to the eigenfunctions of the eigenmode specified by $n$ and $l$ (\citealt{unno1989}; \citealt{aerts2010}).  For later use, we explicitly introduce the average rotation rate, $\bar{\Omega}_{n,l}$, by
\begin{equation}
\bar{\Omega}_{n,l}
=
 \int_0^R K_{n,l}(r)\Omega(r) dr
.
\label{eq:Omega_average}
\end{equation}
{\it It should be stressed that, although the Ledoux constant, $C_{n,l}$, was originally introduced in the case of uniform rotation, equation\,(\ref{eq:split}) is valid for any rotation profile that depends only on the radius.} We therefore do not need to assume uniform rotation at all in the following analysis, which is based on $C_{n,l}$.

\begin{figure}
\centering
\includegraphics[width=0.98\linewidth,angle=0]{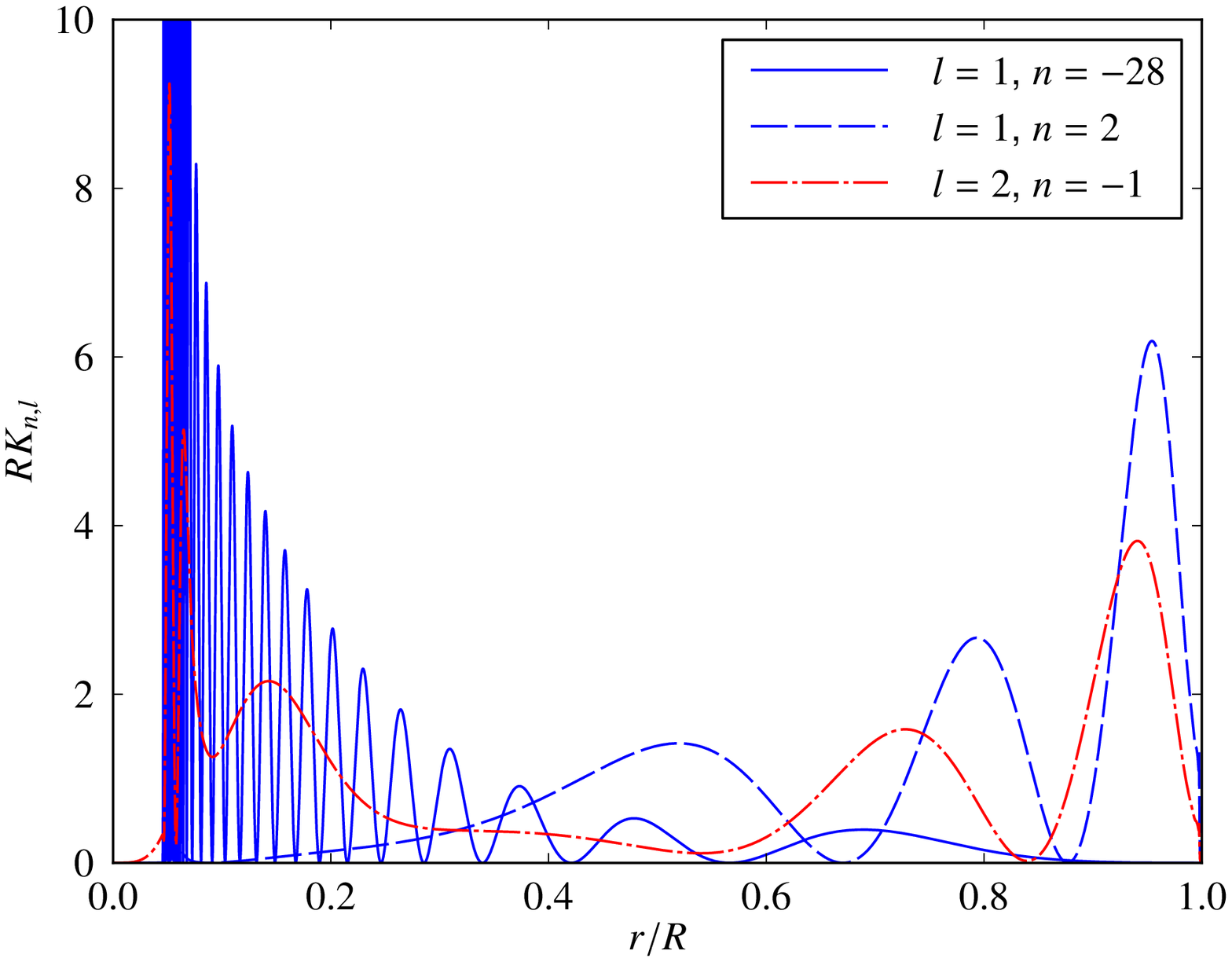}
\\
\includegraphics[width=0.98\linewidth,angle=0,clip=true]{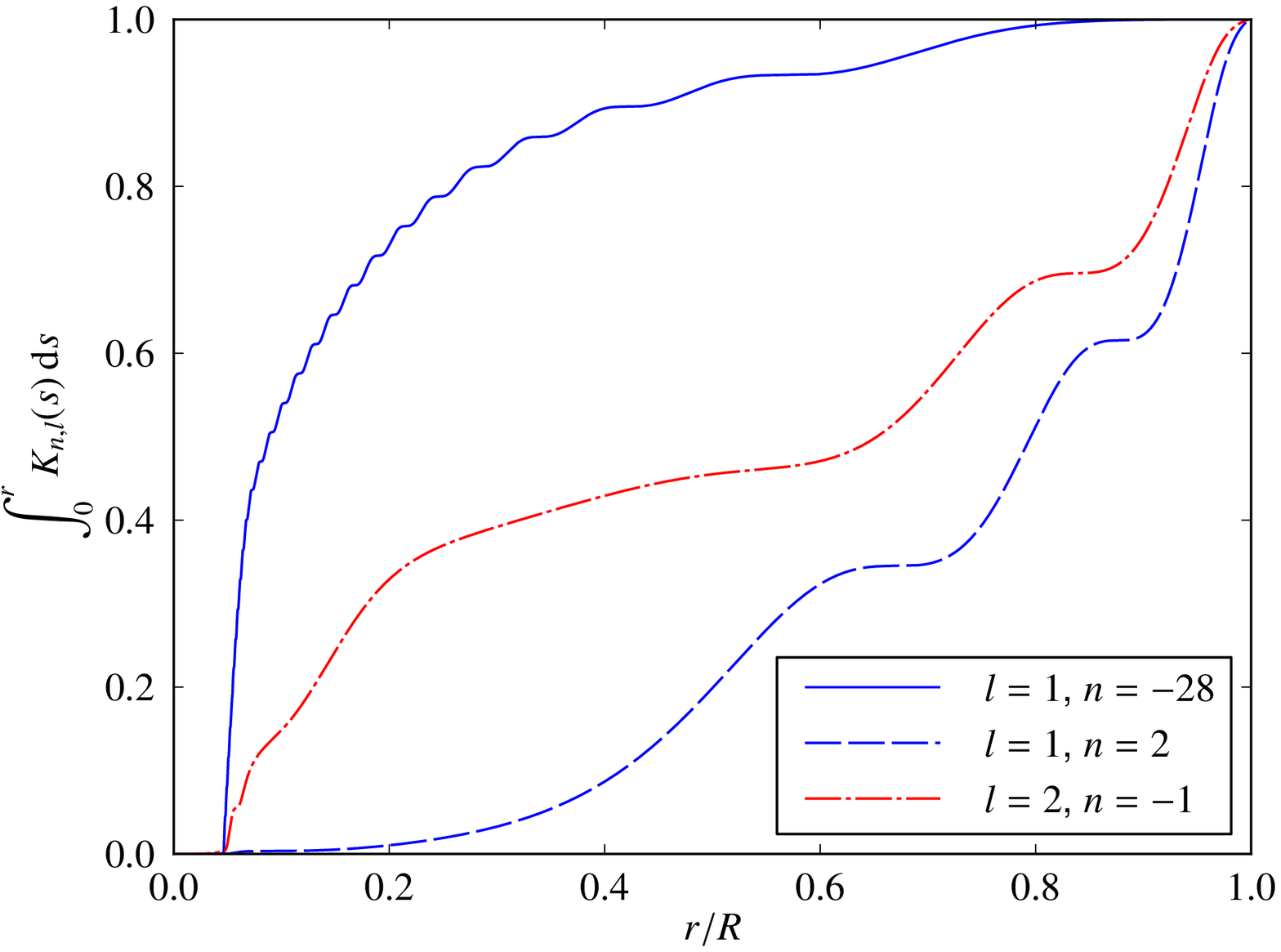}
\caption{Rotation kernels of our best model with 1.46\,M$_{\odot}$ as functions of fractional radius (upper panel) and their cumulative profiles (lower panel). It is clear that the g\,mode (with $l=1$ and $n=-28$) is strongly weighted to the interior 10\,per\,cent in radius (30\,per\,cent in mass), and the dipolar p\,mode (with $l=1$ and $n=2$) is strongly weighted to the outer envelope. The quadrupolar mode (with $l=2$ and $n=-1$) is sensitive to both the core and the envelope, because it is a mixed mode. The g\,mode does not sense the inner few per cent of the core because the core is convective. The g\,mode and the quadrupolar mode are strongly trapped by the steep gradient of the distribution of mean molecular weight, which is located in the range, $0.045 \le r/R \le 0.075$ (see Fig.\,\ref{fig:bvf_xh}).}
\label{fig:kernel}
\end{figure}

Fig.\,\ref{fig:kernel} shows how the observed rotational splittings are weighted in the interior of one of our best models of KIC\,11145123 with 1.46\,M$_{\odot}$. The rotation in the inner core $r \le 0.1\,R$ ($M_r \le 0.30\,M$; where $R$ and $M$ are the total radius and total mass of the star, respectively) dominates the g\,mode. The dipolar p\,mode ($l=1$) has a broader sensitivity in fractional radius, but is strongly confined in fractional mass to the outer 5\,per\,cent of the star. Thus the separation of the dipolar p\,modes and g\,modes in frequency in KIC\,11145123 measures both surface and core rotation rates nearly independently. On the other hand, the quadrupolar mode ($l=2$) has almost the equal sensitivity in the core and the envelope, because it has mixed characters of acoustic waves in the envelope and gravity waves in the core.

\subsection{Nearly model-independent inferences}
\label{subset:model-indept_inf} 

We measured the frequency splittings for the 15 g-mode dipole triplets, and for a selection of p-mode triplets and quintuplets, as seen in Tables\,\ref{table:11145123_g} and \ref{table:11145123_p}. For all of the p-mode multiplets, the splittings within the multiplet are equal within the formal errors. This shows that there are no second-order rotation effects measureable and that there is no strong magnetic field in the star, as that would perturb the multiplet frequency spacings. The equally split quintuplets indicate no detectable differential rotation in the latitudinal direction.

For high-overtone g\,modes the Ledoux constant asymptotically approaches $C_{n,l} \approx 1/l(l+1)$, which for the $l = 1$ dipole g\,modes in KIC\,11145123 gives $C_{n,l} \approx 0.5$. This is a very general conclusion, with which our best models constructed in section~\ref{sec:model} are certainly consistent. Using equation\,(\ref{eq:split}) for the g-mode rotational triplets, we find $P_{\rm rot} {\rm (core)} = 104 - 105$\,d. On the other hand, using a typical value of the Ledoux constant for the p\,modes of our models, $C_{n,l} = 0.03$, the p-mode triplets give surface rotation periods in the range $95 - 114$\,d. Although the above estimates do not take account of errors in $C_{n,l}$, the consistent period range of the p\,modes with that of the g\,modes supports uniform rotation of the star in a broad way.

We can even show based on a careful argument about the average rotation rate, $\bar{\Omega}$, that the star is rotating differentially in the sense that the envelope rotation rate is slightly {\it higher} than the core rate.

Because this is a very important conclusion,
we present a careful 4-step argument,
whose essential points are summarised before going into the details.
First of all, it is intended to demonstrate that
the average rotation rate of the envelope layers probed by a p mode
is higher than that of the core layers inferred by a g mode.
We first show that the rotational splitting of the p mode is 
more than twice as large as that of the g mode.
Then, we argue based on stellar pulsation theory that
twice the g-mode splitting provides the upper limit of
the core rotation rate,
and that the p-mode splitting constrains the lower limit of
the envelope rate.
Combining these points, we are led to the conclusion.
Now the details of the argument follow.
\begin{enumerate}
 \item
The frequency separations of the g\,modes are not exactly half of those of the p\,modes, as has already been pointed out in section \ref{sec:frequencies}. In fact, if we pay attention to the p- and g-mode triplets with the highest amplitudes (those centred on $f = 1.418\,{\rm d}^{-1}$  in Table \ref{table:11145123_p} and $f = 18.366\,{\rm d}^{-1}$ in Table \ref{table:11145123_g}, respectively),  whose frequencies are determined most precisely, their splittings are given by
\begin{equation}
\Delta f({\rm p})= 0.0101453 \pm 0.0000023\,{\rm d}^{-1}
\end{equation}
and
\begin{equation}
\Delta f({\rm g}) = 0.0047562 \pm 0.0000010\,{\rm d}^{-1}\;,
\end{equation}
giving 
\begin{equation}
\Delta f({\rm p}) - 2\Delta f({\rm g}) = 0.0006329 \pm 0.0000030\,{\rm d}^{-1}\;.
\end{equation}
Note that we generally put $({\rm p})$ and $({\rm g})$ to quantities that are associated with these (best-measured) dipolar p and g\,modes in this subsection. Since $\Delta f = \delta\omega/(2\pi)$, it can be claimed with strong statistical significance that
\begin{equation}
 \delta\omega({\rm p}) > 2 \delta\omega({\rm g})
\;.
\label{eq:faster_env_1}
\end{equation}

\item
The Ledoux constant, $C_{n,l}$, of high-order dipolar g\,modes is not exactly equal to $\frac{1}{2}$, but is a little smaller. This is because a detailed asymptotic analysis of high-order dipolar g\,modes shows that  $C_{n,1}$ approaches $\frac{1}{2}$ from below [see equations\,(\ref{eq:Cn1_upper_limit}) and (\ref{eq:Cn1_limit}) in appendix \ref{appendix:Cn1}].
A consequence of $C_{n,1} < \frac{1}{2}$ with equations\,(\ref{eq:split}) and (\ref{eq:Omega_average}) is that
\begin{equation}
2 \delta \omega({\rm g}) = 2 ( 1 - C_{n,1} )
\bar{\Omega}({\rm g}) > \bar{\Omega}({\rm g})
.
\label{eq:faster_env_2}
\end{equation}
Namely, the upper limit of $\bar{\Omega}({\rm g})$ is given by $2 \delta \omega({\rm g})$. The corresponding lower limit of the average rotation period is given by $105.13\pm 0.02$\,d.

\item
\label{item:positive_Cnl}
If $C_{n,l} > 0$ for the p\,mode, the rotation rate in the envelope is constrained from below. Assuming $C_{n,l} > 0$ in equation\,(\ref{eq:split}), it is found
\begin{equation}
\delta\omega({\rm p}) = ( 1 - C_{n,l} ) \bar{\Omega}({\rm p})
< \bar{\Omega}({\rm p})
.
\label{eq:faster_env_3}
\end{equation}
Therefore, the lower limit of $\bar{\Omega}({\rm p})$ is provided by $\delta\omega({\rm p})$. The corresponding upper limit of the average rotation period is given by $98.57 \pm 0.02$\,d.

\item
Equations
(\ref{eq:faster_env_1}),
(\ref{eq:faster_env_2}) and
(\ref{eq:faster_env_3})
lead to
\begin{equation}
\bar{\Omega}({\rm p}) > \bar{\Omega}({\rm g})
,
\label{eq:Omega_p_g_inequality}
\end{equation}
which implies the rotation rate of envelope layers that are probed by the p\,modes are on average higher than that of core layers that are diagnosed by the g\,modes. The corresponding average rotation period of the envelope layers is at least 7\,per\,cent shorter than that of the core layers.

\end{enumerate}

Some comments about the crucial assumption, $C_{n,l} > 0$, at step \ref{item:positive_Cnl} above follow. Although it is generally possible to find an eigenmode with $C_{n,l} < 0$, such modes seem to be rare \citep[e.g.][]{gough2002}. In fact, not a single mode with $C_{n,l} < 0$ has been found in any of our evolutionary models in section \ref{sec:model}, even if  those that cannot reproduce the observed frequencies are included. On the other hand, we have confirmed that some low-order dipolar p\,modes of polytropic models with index higher than $3.9$ have $C_{n,l} < 0$. However, the values are no less than about $-0.002$, whereas a value of $C_{n,l} \le -0.07$ would be required to conclude $\bar{\Omega}({\rm p}) \le \bar{\Omega}({\rm g})$.

{\it We stress once again that the above argument is based on only conservative assumptions that are not influenced by detailed modelling of the star and precise mode identification.} For example, one of our fundamental assumptions is that the p\,mode is more sensitive to outer layers of the star than the g\,mode. This is generally true for any pair of a p\,mode and a high-order g\,mode in any main-sequence star. Moreover, although we rely on the identification of the g\,mode  as the one with $l=1$ and  a large radial order ($|n| \gg 1$),  the exact value of $n$ need not be specified, as is the case for the p\,mode. Therefore, our conclusion of the higher rotation rate in the envelope than in the core is robust.

\subsection{Two-zone modelling}
\label{sec:two-zone_modelling}

The robust conclusion obtained in the last subsection was made possible by the structure of the rotation kernels. Independently of the details of the model, the g-mode kernels are primarily confined in the core, whereas the p-mode kernels have large amplitudes only in the outer layers. The situation also permits the following simplified two-zone modelling.

In constructing a two-zone model, in which the angular velocity $\Omega(r)$ has the form
\begin{equation}
\Omega(r)=\left\{
\begin{array}{ll}
\Omega_1 & (0\le r\le r_{\rm b}) \\
\Omega_2 & (r_{\rm b}\le r \le R)
\end{array}
\right.
\label{eq:two-zone_model}
\end{equation}
with a prescribed position of the boundary $r_{\rm b}$, one usually fits observed rotational shifts of frequencies. Here, however, because of the segregation of their kernels, we can simply average g-mode data and p-mode data separately and use them to obtain a two-zone model. Namely, we define the mean rotational shift for g\,modes as
\begin{equation}
\delta\omega_{\rm g} = \sum_{\rm g}c_{{\rm g},nlm}\delta\omega_{n,l,m}~,
\end{equation}
where the summation is taken over the g\,modes we use for the analysis. The weighting coefficients $c_{{\rm g},nlm}$ are inversely proportional to the formal uncertainty in $\delta\omega_{n,l,m}$, and the sum of all the coefficients is unity. Using the same weighting coefficients, observation errors are propagated to $\delta\omega_{\rm g}$, and we also introduce a mean splitting kernel for g\,modes, including the factor $m(1-C_{n,l})$:
\begin{equation}
K_{\rm g}(r) = \sum_{\rm g}c_{{\rm g},nlm}m(1-C_{n,l})K_{n,l}(r)
\end{equation}
to have the g-mode constraint
\begin{equation}
\delta\omega_{\rm g} = \int_0^R K_{\rm g}(r) \Omega(r) dr \; ,
\end{equation}
which is further simplified, assuming the two-zone model (\ref{eq:two-zone_model}), to
\begin{equation}
\delta\omega_{\rm g} = \gamma_{{\rm g}1}\Omega_1 + \gamma_{{\rm g}2}\Omega_2
\label{eq:g-mode_const}
\end{equation}
by denoting two kernel integrals with $\gamma_{{\rm g}i}$:
\begin{eqnarray}
\gamma_{{\rm g}1} &=& \int_0^{r_{\rm b}} K_{\rm g}(r) dr\\
\gamma_{{\rm g}2} &=& \int_{r_{\rm b}}^R K_{\rm g}(r) dr~.
\end{eqnarray}

We repeat the process for p\,modes to obtain the p-mode constraint \begin{equation}
\delta\omega_{\rm p} = \gamma_{{\rm p}1}\Omega_1 + \gamma_{{\rm p}2}\Omega_2 \; ,
\label{eq:p-mode_const}
\end{equation}
with p-mode quantities defined similarly. Note that, in reality, we used $(\omega_{n, l, +1}-\omega_{n, l, -1})/2$ as the rotational-shift measurement for each triplet, with formal errors properly propagated. It is then a straightforward and transparent process to determine $\Omega_1$ and $\Omega_2$ from equations\,(\ref{eq:g-mode_const}) and (\ref{eq:p-mode_const}) and estimate the formal errors.

\begin{figure}
\centering
\includegraphics[width=0.98\linewidth,angle=0]{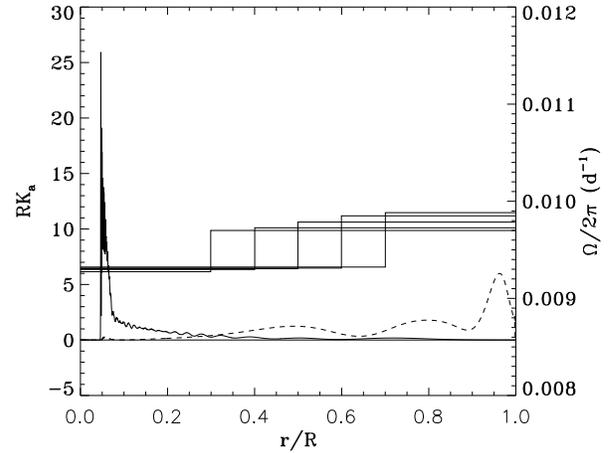}
\caption{The step functions are the rotational frequencies ($\Omega(r)/2\pi$) of the two-zone model obtained from 15 g\,modes and 2 dipole p\,modes, for the boundary position $r_{\rm b}/R=0.3, 0.4, 0.5, 0.6$ and $0.7$. The g-mode kernel $K_{\rm g}(r)$ (solid curve) and the p-mode kernel $K_{\rm p}(r)$ (dashed curve) are also shown to be well separated. The kernels are multiplied by the stellar radius $R$ to render them dimensionless.
\label{fig:two-zone_models}
}
\end{figure}

Fig.\,\ref{fig:two-zone_models} shows the results obtained by using all 15 g\,modes and two dipole p\,modes, $\nu_2$ and $\nu_5$, all given in Table\,\ref{table:mode_id}, together with mean splitting kernels for g and p\,modes, which clearly show that the spatial separation for these mean kernels is indeed good. We have good measurements of two quadrupole mixed modes ($\nu_3$ and $\nu_4$ in Table\,\ref{table:mode_id}), but they are excluded from this simple analysis because their splitting kernels have amplitudes both in the core and in the outer layers. This is in contrast to the situation in the subgiant and giant stars studied by \citet{deheuvels2014} and by \citet{beck2012}, where they had to rely on mixed modes to probe the deep interior.

The most natural choice of the boundary is $r_{\rm b} = 0.3R$, as this is where the g-mode kernel and the p-mode kernel cross over, but results for other values of $r_{\rm b}$ are also shown. For $r_{\rm b}=0.3R$ the estimates obtained are $\Omega_1/2\pi = 0.009379\pm0.000002$\,(d$^{-1}$) and $\Omega_2/2\pi = 0.009695\pm0.000002$\,(d$^{-1}$). As we see in Fig. \ref{fig:two-zone_models}, the estimates do not depend strongly on the choice of $r_{\rm b}$. Essentially, these results strengthen the inference that in this star the surface layer is rotating slightly faster than the core. The surface to core contrast of the rotation speed is about 3 per cent, less than was indicated in Section\,\ref{subset:model-indept_inf}. This is due to the inclusion of $\nu_5$, that exhibits rotational shifts of frequencies much less than $\Delta f({\rm p})$ in equation\,(8).

\begin{figure}
\centering
\includegraphics[width=0.98\linewidth,angle=0]{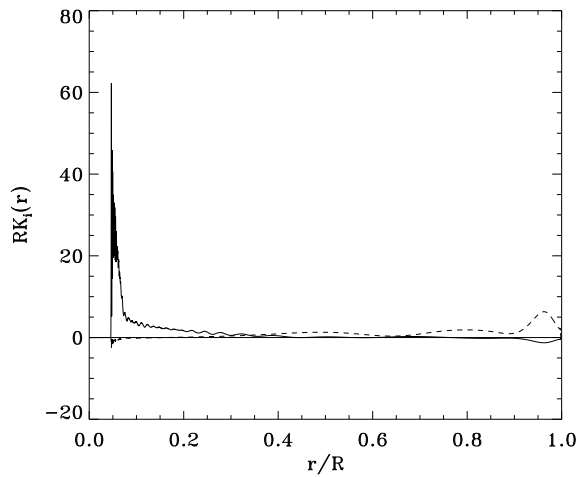}  
\caption{Averaging kernels $K_1(r)$ (solid) and $K_2(r)$ (dashed) for the $r_{\rm b}=0.3 R$ case, with the thin horizontal line to indicate the zero level. It is seen that $\Omega_1$ and $\Omega_2$ indeed represent the core rotation rate and the surface rate well.
The kernels are multiplied by the stellar radius $R$ to render them dimensionless.
\label{fig:two-zone-model_avkrns}
}
\end{figure}

Since our estimates of $\Omega_i$ are simply linear combinations of $\delta\omega_{\rm g}$ and $\delta\omega_{\rm p}$, which are in turn linear combinations of $\delta\omega_{n,l,m}$ in  equation\,(\ref{eq:split}), it is possible to write down  the relation between acquired $\Omega_i$ and the rotation rate in the star in the following form
\begin{equation}
\Omega_i = \int_0^R K_i(r)\Omega(r)dr~.
\end{equation}
The unimodular function $K_i(r)$ is the averaging kernel, used to examine the resolution or, in our case, unwanted contamination by rotation rate outside the target region. For example, if $K_1(r)$ has a large value near the surface, it means our estimate of the `core rotation rate' is affected by the surface rate. In our case, integrating $K_1(r)$ in the range $0\le r\le r_{\rm b}$ and integrating $K_2(r)$ in the range $r_{\rm b}\le r\le R$ return unity. As is seen in Fig. \ref{fig:two-zone-model_avkrns}, the averaging kernel $K_1(r)$ is not completely zero in the range $r_{\rm b}\le r\le R$, nor is $K_2(r)$ in the range $0\le r\le r_{\rm b}$, but after integration they are essentially zero, giving nearly complete separation of the zones. The small contribution from the off-target ranges are the indication that our choice of $r_{\rm b}=0.3R$ was a good one.

\subsubsection{Inversion}
\label{sec:inversion}

Regularised Least-Squares fitting with a first derivative constraint has also been done, using 15 g\,modes and 4 p\,modes ($\nu_2$ to $\nu_5$ in Table\,\ref{table:mode_id}). For strong regularisation parameters (i.e. strong smoothing), once again we obtain rotation rates that indicate that the surface layer is rotating slightly faster than the core. 

\section{Discussion and conclusions}
\label{sec:discussion}

These results are remarkable for several reasons. The most important of these is that we see surface-to-core rotation clearly for the first time in a star burning hydrogen in the core. Secondly, the star is nearly a rigid rotator, but the surface layer for the highest amplitude p\,mode rotates more quickly than does the core. This is unexpected. Thirdly, the 100-d rotation period of KIC\,11145123 is abnormally long for any nonmagnetic A star. 

Amongst the magnetic Ap stars rotation periods of years are known; the longest is over a century for $\gamma$\,Equ. However, we consider it unlikely that KIC\,11145123  is currently a magnetic Ap star: the equal splitting of the p-mode multiplets argues against a magnetic field. No magnetic Ap star is known to pulsate in low-overtone p\,modes, or in g\,modes. While there is a class of rapidly oscillating Ap stars \citep{kurtz1990} that pulsate in high-overtone p\,modes, KIC\,11145123 does not show any of these, and \citet{saio2005} has shown that the strong magnetic fields of the Ap stars suppress the low-overtone p\,modes typical of $\delta$\,Sct stars. All of these reasons argue against a magnetic field in KIC\,11145123.

The high helium abundances of our best models suggest that KIC\,11145123 could be an SX\,Phe variable that was formed in a binary system after a significant mass accretion. If this is the case, then the envelope may have been spun up considerably. In this scenario, at present, near the end of main sequence evolution, only a slight excess of angular frequency in the envelope remains and the rotation is nearly uniform and very slow. The small difference in the
rotation rate between the core and the envelope suggests that the angular momentum transfer may be much stronger than has been previously thought. This is common to the requirement suggested from the recent discoveries that the rotation speeds of red giant cores, which are much slower than the theoretical predictions. High-resolution spectra will test this idea by determining abundances of, in particular, CNO, and will test our prediction above that KIC\,11145123 is not a magnetic Ap star.

While the 100-d rotation period for KIC\,11145123 is long for an A star, we note that \citet{aerts2003} found a similarly long 80-d rotation period in the early B star HD\,129929 for which they found evidence of internal differential rotation. In the case of KIC\,11145123 it is the slow rotation that makes the rotational splitting patterns so obvious. There are many hybrid $\delta$\,Sct -- $\gamma$\,Dor stars in the {\it Kepler} data that show an abundance of g\,modes and p\,modes. But with the typically much faster rotation rate than for KIC\,11145123, it is probable that the rotationally split multiplets for these stars are not equally split, because of second order effects. Then with the richness of the g-mode frequency spectrum, it is difficult to determine which of the plethora of frequency peaks in the amplitude spectrum belong to dipole or quadrupole multiplets in these faster rotators, since the multiplet patterns get mixed in frequency space. Progress may be made by understanding the slower rotators such as KIC\,11145123 first, allowing us then to begin to understand the frequency patterns of more and more rapidly rotating stars. Studies of these stars benefit from the long 4-yr time-span of the {\it Kepler} data, since the mode frequency separations for the g\,modes are small. 

Asteroseismology has revealed internal differential rotation in the outer half of the Sun \citep{schou1998}, has put constraints on interior rotation of some main sequence B stars (\citealt{aerts2003}; \citealt{pamyatnykh2004}; \citealt{dziembowski2008}; \citealt{briquet2007}), surface-to-core differential rotation in two red giants stars (\citealt{beck2012}; \citealt{deheuvels2012}), and some subgiants \citep{deheuvels2014} and both rigid rotation \citep{charpinet2009} and differential rotation \citep{corsico2011} in two white dwarf stars. 

Stars spend 90\,per\,cent of their lifetimes as main sequence stars. Now, for the first time, we have measured the rotation of a main sequence A star, KIC\,11145123, at the surface and in the core, {\it essentially model-independently}, and with a clarity never seen before. We have found it to be nearly a rigid rotator with the surface rotating slightly faster than the core. With this discovery, rotation and angular momentum transfer inside of main sequence stars is now an observational science. Our understanding of KIC\,11145123 also shows the direction for finding more main sequence stars with similarly rich frequency spectra with many rotational multiplets for both p\,modes and g\,modes.

Stars are born in the wholly convective phase (the Hayashi phase), where uniform rotation is established because of large turbulent viscosity. Subsequent stellar evolution is generally a process of increasing the central mass concentration.  A simple argument based on conservation of the local angular momentum then leads to the conclusion of  a greater rotation rate in the central regions than in the envelope as a result of evolution. Following this scenario, we have modelled the evolution of angular frequency, $\Omega(M_r,t)$, for a hypothetical star similar to our best model for KIC\,11145123. We started with a uniformly rotating, fully convective protostar in its Hayashi phase and evolved it to the TAMS. In this model angular momentum is conserved in radiative layers, and in convective zones; initial uniform rotation was assumed and the total angular momentum was conserved. As the main sequence evolution proceeds, a steep variation in $\Omega$ appears around the core. From the rotation kernels seen in Fig. 4 we conclude that we would have found differential rotation from core to surface in KIC11145123 of a factor of 5, had the star been born with uniform rotation and had angular momentum been conserved. Clearly, a strong mechanism for angular momentum transport must be acting to result in the nearly rigid rotation that we observe.

That the envelope of KIC\,11145123 rotates more rapidly than the core puts a constraint on physical mechanisms of angular momentum transport. We may classify the transport mechanisms into two categories. In the first, the mechanisms operate only to reduce the gradient of the rotation rate. Viscosity (of any kind) is a representative example. On the other hand, the mechanisms in the second category can even reverse the sign of the gradient. Examples are the angular momentum transport by waves and mass accretion (including capture of planets and/or comets). Given the spin-up tendency of the core as a result of evolution, the mechanisms in the first category are clearly not sufficient to explain the more rapidly rotating envelope. We thus conclude that those in the second category must exist.

\section*{acknowledgements}

We thank NASA and the {\it Kepler} team for their revolutionary data. This work was carried out with support from a JSPS Japan-UK Joint Research grant. D. Kurtz thanks the JSPS for a Furusato Award that partially funded this work. H. Saio thanks Bill Paxton for his help in extracting structure data from the MESA code. We thank Steve Kawaler and Tim Bedding for helpful discussions.

\bibliography{11145123_submitted.bib}

\begin{thebibliography}{33}
\expandafter\ifx\csname natexlab\endcsname\relax\def\natexlab#1{#1}\fi

\bibitem[{{Aerts}, {Christensen-Dalsgaard} \& {Kurtz}(2010){Aerts},
  {Christensen-Dalsgaard}, \& {Kurtz}}]{aerts2010}
{Aerts} C., {Christensen-Dalsgaard} J., {Kurtz} D.~W., 2010, {Asteroseismology}

\bibitem[{{Aerts} {et~al}\mbox{.}(2003){Aerts}, {Thoul}, {Daszy{\'n}ska},
  {Scuflaire}, {Waelkens}, {Dupret}, {Niemczura}, \& {Noels}}]{aerts2003}
{Aerts} C., {Thoul} A., {Daszy{\'n}ska} J., {Scuflaire} R., {Waelkens} C.,
  {Dupret} M.~A., {Niemczura} E., {Noels} A., 2003, Science, 300, 1926

\bibitem[{{Appourchaux} {et~al}\mbox{.}(2010){Appourchaux}, {Belkacem},
  {Broomhall}, {Chaplin}, {Gough}, {Houdek}, {Provost}, {Baudin}, {Boumier},
  {Elsworth}, {Garc{\'{\i}}a}, {Andersen}, {Finsterle}, {Fr{\"o}hlich},
  {Gabriel}, {Grec}, {Jim{\'e}nez}, {Kosovichev}, {Sekii}, {Toutain}, \&
  {Turck-Chi{\`e}ze}}]{appourchaux2010}
{Appourchaux} T. {et~al.}, 2010, \aapr, 18, 197

\bibitem[{{Asplund} {et~al}\mbox{.}(2009){Asplund}, {Grevesse}, {Sauval}, \&
  {Scott}}]{asplund2009}
{Asplund} M., {Grevesse} N., {Sauval} A.~J., {Scott} P., 2009, \araa, 47, 481

\bibitem[{{Beck} {et~al}\mbox{.}(2012){Beck}, {Montalban}, {Kallinger}, {De
  Ridder}, {Aerts}, {Garc{\'{\i}}a}, {Hekker}, {Dupret}, {Mosser},
  {Eggenberger}, {Stello}, {Elsworth}, {Frandsen}, {Carrier}, {Hillen},
  {Gruberbauer}, {Christensen-Dalsgaard}, {Miglio}, {Valentini}, {Bedding},
  {Kjeldsen}, {Girouard}, {Hall}, \& {Ibrahim}}]{beck2012}
{Beck} P.~G. {et~al.}, 2012, \nat, 481, 55

\bibitem[{{Briquet} {et~al}\mbox{.}(2007){Briquet}, {Morel}, {Thoul},
  {Scuflaire}, {Miglio}, {Montalb{\'a}n}, {Dupret}, \& {Aerts}}]{briquet2007}
{Briquet} M., {Morel} T., {Thoul} A., {Scuflaire} R., {Miglio} A.,
  {Montalb{\'a}n} J., {Dupret} M.-A., {Aerts} C., 2007, \mnras, 381, 1482

\bibitem[{{Charpinet}, {Fontaine} \& {Brassard}(2009){Charpinet}, {Fontaine},
  \& {Brassard}}]{charpinet2009}
{Charpinet} S., {Fontaine} G., {Brassard} P., 2009, \nat, 461, 501

\bibitem[{{C{\'o}rsico} {et~al}\mbox{.}(2011){C{\'o}rsico}, {Althaus},
  {Kawaler}, {Miller Bertolami}, {Garc{\'{\i}}a-Berro}, \&
  {Kepler}}]{corsico2011}
{C{\'o}rsico} A.~H., {Althaus} L.~G., {Kawaler} S.~D., {Miller Bertolami}
  M.~M., {Garc{\'{\i}}a-Berro} E., {Kepler} S.~O., 2011, \mnras, 418, 2519

\bibitem[{{Cowling}(1941)}]{cowling1941}
{Cowling} T.~G., 1941, \mnras, 101, 367

\bibitem[{{Deheuvels} {et~al}\mbox{.}(2014){Deheuvels}, {Do{\u g}an}, {Goupil},
  {Appourchaux}, {Benomar}, {Bruntt}, {Campante}, {Casagrande}, {Ceillier},
  {Davies}, {De Cat}, {Fu}, {Garc{\'{\i}}a}, {Lobel}, {Mosser}, {Reese},
  {Regulo}, {Schou}, {Stahn}, {Thygesen}, {Yang}, {Chaplin},
  {Christensen-Dalsgaard}, {Eggenberger}, {Gizon}, {Mathis},
  {Molenda-{\.Z}akowicz}, \& {Pinsonneault}}]{deheuvels2014}
{Deheuvels} S. {et~al.}, 2014, ArXiv e-prints

\bibitem[{{Deheuvels} {et~al}\mbox{.}(2012){Deheuvels}, {Garc{\'{\i}}a},
  {Chaplin}, {Basu}, {Antia}, {Appourchaux}, {Benomar}, {Davies}, {Elsworth},
  {Gizon}, {Goupil}, {Reese}, {Regulo}, {Schou}, {Stahn}, {Casagrande},
  {Christensen-Dalsgaard}, {Fischer}, {Hekker}, {Kjeldsen}, {Mathur}, {Mosser},
  {Pinsonneault}, {Valenti}, {Christiansen}, {Kinemuchi}, \&
  {Mullally}}]{deheuvels2012}
---, 2012, \apj, 756, 19

\bibitem[{{Dupret} {et~al}\mbox{.}(2004){Dupret}, {Thoul}, {Scuflaire},
  {Daszy{\'n}ska-Daszkiewicz}, {Aerts}, {Bourge}, {Waelkens}, \&
  {Noels}}]{dupret2004}
{Dupret} M.-A., {Thoul} A., {Scuflaire} R., {Daszy{\'n}ska-Daszkiewicz} J.,
  {Aerts} C., {Bourge} P.-O., {Waelkens} C., {Noels} A., 2004, \aap, 415, 251

\bibitem[{{Dziembowski} \& {Pamyatnykh}(2008)}]{dziembowski2008}
{Dziembowski} W.~A., {Pamyatnykh} A.~A., 2008, \mnras, 385, 2061

\bibitem[{{Eddington}(1926)}]{eddington1926}
{Eddington} A.~S., 1926, {The Internal Constitution of the Stars}

\bibitem[{{Gough}(2002)}]{gough2002}
{Gough} D.~O., 2002, in Astronomical Society of the Pacific Conference Series,
  Vol. 259, IAU Colloq. 185: Radial and Nonradial Pulsationsn as Probes of
  Stellar Physics, {Aerts} C., {Bedding} T.~R., {Christensen-Dalsgaard} J.,
  eds., p.~37

\bibitem[{{Huber} {et~al}\mbox{.}(2014){Huber}, {Silva Aguirre}, {Matthews},
  {Pinsonneault}, {Gaidos}, {Garc{\'{\i}}a}, {Hekker}, {Mathur}, {Mosser},
  {Torres}, {Bastien}, {Basu}, {Bedding}, {Chaplin}, {Demory}, {Fleming},
  {Guo}, {Mann}, {Rowe}, {Serenelli}, {Smith}, \& {Stello}}]{huber2014}
{Huber} D. {et~al.}, 2014, \apjs, 211, 2

\bibitem[{{Iglesias} \& {Rogers}(1996)}]{opal}
{Iglesias} C.~A., {Rogers} F.~J., 1996, \apj, 464, 943

\bibitem[{{Kurtz}(1990)}]{kurtz1990}
{Kurtz} D.~W., 1990, \araa, 28, 607

\bibitem[{{Ledoux}(1951)}]{ledoux1951}
{Ledoux} P., 1951, \apj, 114, 373

\bibitem[{{Leighton}, {Noyes} \& {Simon}(1962){Leighton}, {Noyes}, \&
  {Simon}}]{leighton1962}
{Leighton} R.~B., {Noyes} R.~W., {Simon} G.~W., 1962, \apj, 135, 474

\bibitem[{{Miglio} {et~al}\mbox{.}(2008){Miglio}, {Montalb\'an}, {Noels}, \&
  {Eggenberger}}]{miglio2008}
{Miglio} A., {Montalb\'an} J., {Noels} A., {Eggenberger} P., 2008, 386, 1487

\bibitem[{{Pamyatnykh}, {Handler} \& {Dziembowski}(2004){Pamyatnykh},
  {Handler}, \& {Dziembowski}}]{pamyatnykh2004}
{Pamyatnykh} A.~A., {Handler} G., {Dziembowski} W.~A., 2004, \mnras, 350, 1022

\bibitem[{{Paxton} {et~al}\mbox{.}(2013){Paxton}, {Cantiello}, {Arras},
  {Bildsten}, {Brown}, {Dotter}, {Mankovich}, {Montgomery}, {Stello}, {Timmes},
  \& {Townsend}}]{paxton2013}
{Paxton} B. {et~al.}, 2013, \apjs, 208, 4

\bibitem[{{Pinsonneault}(1997)}]{pinsonneault1997}
{Pinsonneault} M., 1997, \araa, 35, 557

\bibitem[{{Saio}(2005)}]{saio2005}
{Saio} H., 2005, \mnras, 360, 1022

\bibitem[{{Saio} \& {Cox}(1980)}]{saio1980}
{Saio} H., {Cox} J.~P., 1980, \apj, 236, 549

\bibitem[{{Schou} {et~al}\mbox{.}(1998){Schou}, {Antia}, {Basu}, {Bogart},
  {Bush}, {Chitre}, {Christensen-Dalsgaard}, {di Mauro}, {Dziembowski},
  {Eff-Darwich}, {Gough}, {Haber}, {Hoeksema}, {Howe}, {Korzennik},
  {Kosovichev}, {Larsen}, {Pijpers}, {Scherrer}, {Sekii}, {Tarbell}, {Title},
  {Thompson}, \& {Toomre}}]{schou1998}
{Schou} J. {et~al.}, 1998, \apj, 505, 390

\bibitem[{{Stellingwerf}(1979)}]{stellingwerf1979}
{Stellingwerf} R.~F., 1979, \apj, 227, 935

\bibitem[{{Takata}(2006)}]{takata2006}
{Takata} M., 2006, \pasj, 58, 893

\bibitem[{{Takata}(2012)}]{takata2012}
---, 2012, \pasj, 64, 66

\bibitem[{{Tayar} \& {Pinsonneault}(2013)}]{tayar2013}
{Tayar} J., {Pinsonneault} M.~H., 2013, \apjl, 775, L1

\bibitem[{{Unno} {et~al}\mbox{.}(1989){Unno}, {Osaki}, {Ando}, {Saio}, \&
  {Shibahashi}}]{unno1989}
{Unno} W., {Osaki} Y., {Ando} H., {Saio} H., {Shibahashi} H., 1989, {Nonradial
  oscillations of stars}

\bibitem[{{Uytterhoeven} {et~al}\mbox{.}(2011){Uytterhoeven}, {Moya},
  {Grigahc{\`e}ne}, {Guzik}, {Guti{\'e}rrez-Soto}, {Smalley}, {Handler},
  {Balona}, {Niemczura}, {Fox Machado}, {Benatti}, {Chapellier}, {Tkachenko},
  {Szab{\'o}}, {Su{\'a}rez}, {Ripepi}, {Pascual}, {Mathias},
  {Mart{\'{\i}}n-Ru{\'{\i}}z}, {Lehmann}, {Jackiewicz}, {Hekker},
  {Gruberbauer}, {Garc{\'{\i}}a}, {Dumusque}, {D{\'{\i}}az-Fraile}, {Bradley},
  {Antoci}, {Roth}, {Leroy}, {Murphy}, {De Cat}, {Cuypers}, {Kjeldsen},
  {Christensen-Dalsgaard}, {Breger}, {Pigulski}, {Kiss}, {Still}, {Thompson},
  \& {van Cleve}}]{uytterhoeven2011}
{Uytterhoeven} K. {et~al.}, 2011, \aap, 534, A125

\end{thebibliography}

\appendix
\section{The Ledoux constant of high-order dipolar g\,modes}
\label{appendix:Cn1}

\subsection{The Ledoux constant of dipolar modes}

\citet{takata2006} showed that adiabatic dipolar oscillations of stars can be described by a second-order system of ordinary differential equations, whose dependent variables are defined by
\begin{equation}
 \zeta_r =
J
\xi_r
+
\frac{1}{3g}
\left(
\Phi'
-
r \frac{d\Phi'}{dr}
\right)
\end{equation}
and
\begin{equation}
  \zeta_{\rm h} =
J
\xi_{\rm h}
+
\frac{1}{3 \omega^2 r}
\left[
\left(
1 - 3 J
\right)
\Phi'
-
r
\frac{d\Phi'}{dr}
\right]
.
\end{equation}
Here, the meanings of the symbols are as follows:
$\xi_r$ and $\xi_{\rm h}$
are
the radial part of the radial and horizontal components of the displacement vector, respectively; $\Phi'$ is the Eulerian perturbation to the gravitational potential; $g$ is the gravitational acceleration; $\omega$ is the angular frequency of oscillation; $r$ is the radius; $J$ is defined by 
\begin{equation}
 J = 1 - \frac{4\pi r^3 \rho}{3 M_r}
,
\end{equation}
in which $\rho$ and $M_r$ are the density and the concentric mass, respectively.
A system of differential equations that
$\zeta_r$ and $\zeta_{\rm h}$ satisfy
is provided by
\begin{equation}
 r \frac{d}{dr}
\left(
\begin{array}{c}
 \zeta_r\\
 \zeta_{\rm h}
\end{array}
\right)
=
\left(
\begin{array}{cc}
\frac{V_g}{J} + J - 3 & 2 J - \frac{\lambda V_g}{J}
\\
J - \frac{A^*}{J \lambda} &
\frac{A^*}{J} + 2 J - 3
\end{array}
\right)
\left(
\begin{array}{c}
 \zeta_r\\
 \zeta_{\rm h}
\end{array}
\right)
,
\label{eq:dipolar_system}
\end{equation}
where
$\lambda$, $V_g$ and $A^*$ are defined by
\begin{equation}
 \lambda = \frac{\omega^2 r}{g}
,
\end{equation}
\begin{equation}
 V_g = \frac{g r}{c^2}
\end{equation}
and
\begin{equation}
 A^* = \frac{1}{\Gamma_1} \frac{d\ln p}{d\ln r} - \frac{d\ln\rho}{d\ln r}
,
\end{equation}
respectively.
Here, $\Gamma_1$ is the first adiabatic index, while $p$ is the pressure.

Note that
we may relate $\zeta_r$ and $\zeta_{\rm h}$ to
$\mathcal{Y}_1^{\rm a}$
and
$\mathcal{Y}_2^{\rm a}$,
which are defined by equations (A80) and (A81)
of \citet{takata2006}, respectively.
In fact, there exist simple relations,
$\zeta_r = r \mathcal{Y}_1^{\rm a}$
and
$\zeta_{\rm h} = (r/\lambda) \mathcal{Y}_2^{\rm a}$.
Correspondingly,
equation (\ref{eq:dipolar_system})
can be derived from
equation (A82) of \citet{takata2006},
which is
the system satisfied by
$\mathcal{Y}_1^{\rm a}$ and $\mathcal{Y}_2^{\rm a}$.
The dependent variables $\zeta_r$ and $\zeta_{\rm h}$ are 
related to $\xi_r$ and $\xi_{\rm h}$, respectively, as
\begin{equation}
 \xi_r = \zeta_r + \eta
\end{equation}
and
\begin{equation}
 \xi_{\rm h} = \zeta_{\rm h} + \eta
,
\end{equation}
in which $\eta$ is
defined by
\begin{equation}
\eta
=
-\int_r^R
\frac{4 \pi r^2 \rho}{3 M_r}
\left( \zeta_r + 2 \zeta_{\rm h} \right)
dr
\label{eq:eta_def}
\end{equation}
with $R$ being the total radius of the star. From a physical point of view, $\eta$ is the displacement of the centre of mass of the concentric mass within radius $r$.
Note that $\eta$
is related to $Q_2$, which is introduced by equation (A76) of
\citet{takata2006}, as
$\eta = R Q_2$,
and that
equation (\ref{eq:eta_def}) is equivalent to
equation (A77) of \citet{takata2006}.
Since
$\eta = 0$ at $r = R$,
which is apparent from equation\,(\ref{eq:eta_def}), we obtain
\begin{equation}
\int_0^R
\frac{d}{dr} \left( M_r \eta^2 \right) dr
=
0
,
\end{equation}
from which we find
\begin{equation}
\int_0^R
\eta
\left[
3 \eta
+
2
\left( \zeta_r + 2 \zeta_{\rm h} \right)
\right]
r^2 \rho
dr
=
0
.
\label{eq:eta_integral}
\end{equation}
Here, we have used the mass conservation equation, $dM_r/dr = 4\pi r^2 \rho$. Substituting 
\begin{equation}
\xi_{\rm h} \left( 2 \xi_r + \xi_{\rm h} \right)
=
\zeta_{\rm h} \left( 2 \zeta_r + \zeta_{\rm h} \right)
+
\eta \left[ 3 \eta + 2\left(\zeta_r + 2 \zeta_{\rm h}\right) \right]
\end{equation}
and
\begin{equation}
 \xi_r^2 + 2 \xi_{\rm h}^2
=
 \zeta_r^2 + 2 \zeta_{\rm h}^2
+ \eta \left[ 3 \eta + 2 \left( \zeta_r + 2 \zeta_{\rm h} \right) \right]
,
\end{equation}
into equation\,(\ref{eq:cnl}),
we get, with the help of equation\,(\ref{eq:eta_integral}),
\begin{equation}
 C_{n,\, 1}
=
\frac{\int_0^R
\zeta_{\rm h} \left( 2 \zeta_r + \zeta_{\rm h} \right)
 r^2 \rho
dr}
{\int_0^R
\left( \zeta_r^2 + 2 \zeta_{\rm h}^2 \right)
 r^2 \rho
dr}
.
\label{eq:Cnl_dipole}
\end{equation}
We thus find that $C_{n,1}$ can be estimated by replacing $\xi_r$ and $\xi_{\rm h}$ in equation\,(\ref{eq:cnl}) with $\zeta_r$ and $\zeta_{\rm h}$, respectively.

\subsection{An asymptotic estimate of the Ledoux constant at the low-frequency limit}
\label{appendix:Cn1_low_freq_lim}

We demonstrate that the Ledoux constant of dipolar modes
asymptotically approaches $1/2$ from below as the frequency goes to zero,
without 
neglecting $\Phi'$ (the Cowling approximation; \citealt{cowling1941}).
Because the system given by equation\,(\ref{eq:dipolar_system}) has the same form as the one that is obtained by 
neglecting $\Phi'$
, the asymptotic analysis of equation\,(\ref{eq:dipolar_system}) is totally in parallel with that in the Cowling approximation \citep[e.g.][]{unno1989}. We therefore omit the details here. The asymptotic solutions of equation\,(\ref{eq:dipolar_system}) in the limit of $\omega\rightarrow 0$ are formally given by 
\begin{equation}
\zeta_r = A \cos\Psi
\end{equation}
and
\begin{equation}
\zeta_{\rm h} = \frac{B}{2} \sin\Psi
.
\end{equation}
Here, $\Psi$ is a rapidly changing function of $r$, defined in terms of the Brunt-V\"ais\"al\"a frequency, $N$, by
\begin{equation}
 \Psi = \frac{\sqrt{2}}{\omega} \int^r \frac{N}{r} dr
,
\end{equation}
whereas $A$ and $B$ are functions of $r$ that change slowly. Note that we are not interested in the solutions near the turning points and in the evanescent regions, because they hardly contribute to the integrals in equation\,(\ref{eq:Cnl_dipole}). Because of the relation, 
\begin{equation}
\left(
\zeta_r - 2 \zeta_{\rm h}
\right)^2
=
\frac{1}{2}
\left( A^2 + B^2 \right)
\left\{
1 + \cos \left[ 2 \left(\Psi + \psi_0 \right) \right]
\right\}
,
\end{equation}
where $\psi_0$ is a slowly changing function determined by $A$ and $B$, we obtain 
\begin{equation}
 \int_0^R
\left(
\zeta_r - 2 \zeta_{\rm h}
\right)^2
r^2 \rho dr
\approx
\frac{1}{2}
 \int_P
\left( A^2 + B^2 \right)
r^2 \rho dr
,
\label{eq:asymp_int_1}
\end{equation}
in which we have neglected the integral of a function that changes rapidly around $0$, in addition to the contribution near the turning points and in the evanescent regions. The range of the integral on the right-hand side of equation\,(\ref{eq:asymp_int_1}), which is denoted by $P$, is the propagative region. Similarly, we obtain
\begin{equation}
\int_0^R
 \left( 2 \zeta_{\rm h} \right)^2
r^2 \rho dr
\approx
\frac{1}{2}
\int_P
B^2
r^2 \rho dr
\label{eq:asymp_int_2}
\end{equation}
and
\begin{equation}
\int_0^R
\left(
\zeta_r^2 + 2 \zeta_{\rm h}^2
\right)
r^2 \rho dr
\approx
\frac{1}{2}
\int_P
\left(
A^2 + \frac{B^2}{2}
\right)
r^2 \rho dr
.
\label{eq:asymp_int_3}
\end{equation}
If we note
\begin{equation}
\zeta_r^2 + 2 \zeta_{\rm h}^2
- 2
 \zeta_{\rm h} \left( 2 \zeta_r + \zeta_{\rm h} \right)
=
\left(
\zeta_r - 2 \zeta_{\rm h}
\right)^2
-
\left( 2 \zeta_{\rm h} \right)^2
,
\end{equation}
and utilise equations\,(\ref{eq:asymp_int_1})--(\ref{eq:asymp_int_3}), we eventually find from equation\,(\ref{eq:Cnl_dipole})
\begin{equation}
C_{n,1}
\approx
\frac{1}{2}
-
\frac{\int_P
A^2
r^2 \rho dr}
{\int_P
\left(
A^2 + \frac{B^2}{2}
\right)
r^2 \rho dr}
<
\frac{1}{2}
.
\label{eq:Cn1_upper_limit}
\end{equation}
Moreover, since we can show $A/B \propto \omega$, it is found 
\begin{equation}
\frac{\int_P
A^2
r^2 \rho dr}
{\int_P
B^2
r^2 \rho dr}
\rightarrow 0
\;\mbox{as}\;
\omega \rightarrow 0
,
\end{equation}
which implies
\begin{equation}
 C_{n,1} \rightarrow \frac{1}{2}
\;\mbox{as}\;
\omega \rightarrow 0
.
\label{eq:Cn1_limit}
\end{equation}

\end{document}